\documentclass[a4paper,fleqn]{cas-sc}

\usepackage{amsfonts,amssymb}
\usepackage{graphicx}
\usepackage{caption}
\usepackage[ruled,linesnumbered]{algorithm2e}
\usepackage{amsmath}
\usepackage{bbm}
\usepackage{tabularray}
\usepackage[authoryear,longnamesfirst, numbers]{natbib}
\setcitestyle{numbers,square}

\definecolor{els}{RGB}{0,128,172}
\hypersetup{
  colorlinks=true,
  citecolor=els,
  linkcolor=els,
  urlcolor=els}

\usepackage{soul, color, xcolor}
\soulregister\cite7
\soulregister\ref7 
\usepackage{tcolorbox}
\usepackage{subcaption}
\usepackage[flushleft]{threeparttable}
\usepackage{nomencl}
\usepackage{multicol}
\makenomenclature
\def\tsc#1{\csdef{#1}{\textsc{\lowercase{#1}}\xspace}}
\tsc{WGM}
\tsc{QE}
\tsc{EP}
\tsc{PMS}
\tsc{BEC}
\tsc{DE}

\begin{document}
\let\WriteBookmarks\relax
\def\floatpagepagefraction{1}
\def\textpagefraction{.001}



\title [mode = title]{VEC-Sim: A Simulation Platform for Evaluating Service Caching and Computation Offloading Policies in Vehicular Edge Networks}                      



\author[1]{Fan Wu}[type=editor,
                        style=chinese,
                        ]

\credit{Conceptualization, Methodology, Software, Writing – original draft, Visualization, Formal analysis}

\author[1]{Xiaolong Xu}[style=chinese,orcid=0000-0003-4879-9803]
\cormark[1]
\ead{xlxu@ieee.org}
\credit{Supervision, Writing – review \& editing, Funding acquisition}

\affiliation[1]{organization={School of Software},
    addressline={Nanjing University of Information Science and Technology}, 
    city={Nanjing},
    postcode={210044}, 
    country={China}}

\author[2]{Muhammad Bilal}
\credit{Supervision, Project administration, Writing – review \& editing}

\affiliation[2]{
    organization={School of Computing and Communications},
    addressline={Lancaster University}, 
    city={Lancaster},
    postcode={LA1 4WA}, 
    country={United Kingdom}}

\author[1]{Xiangwei Wang}[style=chinese]
\credit{Data curation, Visualization, Validation}

\author[1]{Hao Cheng}[style=chinese]
\credit{Data curation, Validation}

\author[1]{Siyu Wu}[style=chinese]
\credit{Methodology}


\begin{abstract}
Computer simulation platforms offer an alternative solution by emulating complex systems in a controlled manner. However, existing Edge Computing (EC) simulators, as well as general-purpose vehicular network simulators, are not tailored for VEC and lack dedicated support for modeling the distinct access pattern, entity mobility trajectory and other unique characteristics of VEC networks. To fill this gap, this paper proposes VEC-Sim, a versatile simulation platform for in-depth evaluation and analysis of various service caching and computation offloading policies in VEC networks. VEC-Sim incorporates realistic mechanisms to replicate real-world access patterns, including service feature vector, vehicle mobility modeling, evolving service popularity, new service upload and user preference shifts, etc. Moreover, its modular architecture and extensive Application Programming Interfaces (APIs) allow seamless integration of customized scheduling policies and user-defined metrics. A comprehensive evaluation of VEC-Sim's capabilities is undertaken in comparison to real-world ground truths. Results prove it to be accurate in reproducing classical scheduling algorithms and extremely effective in conducting case studies.
\end{abstract}


\begin{highlights}
    \item We conducted an analysis of vehicular network topologies and application characteristics. Models for communication, computation, and caching were developed to serve as a solid groundwork for subsequent simulator design.
    \item We designed a modular and layered architecture for VEC-Sim. Core entities like vehicles, services and connections were encapsulated in classes with well-defined interfaces. Rich APIs were provided to facilitate extension and customization by operators.
    \item To enhance the realism of the simulation, we proposed mechanisms such as spatial distribution generation, service interest vector, DRL-based vehicle mobility modeling, user preference shift and new service upload. Collectively, these mechanisms empower operators to replicate various complex factors inherent in vehicular network environments.
    \item In-depth experiments were conducted to showcase the capabilities of VEC-Sim, validating its feasibility and effectiveness in modeling VEC systems and performing case studies.
\end{highlights}


\maketitle
\nomenclature{EC}{Edge Computing}
\nomenclature{VEC}{Vehicular Edge Computing}
\nomenclature{CDN}{Content Delivery Network}
\nomenclature{API}{Application Programming Interface}
\nomenclature{ITS}{Intelligent Transportation System}
\nomenclature{IoV}{Internet of Vehicles}
\nomenclature{VCC}{Vehicular Cloud Computing}
\nomenclature{VANET}{Vehicular Ad Hoc Network}
\nomenclature{QoS}{Quality of Service}
\nomenclature{QoE}{Quality of Experience}
\nomenclature{ML}{Machine Learning}
\nomenclature{IoT}{Internet of Things}
\nomenclature{MEC}{Mobile Edge Computing}
\nomenclature{SDV}{Software-Defined Vehicle}
\nomenclature{RSU}{Roadside Unit}
\nomenclature{DRL}{Deep Reinforcement Learning}
\nomenclature{ES}{Edge Server}
\nomenclature{CDC}{Cloud Datacenter}
\nomenclature{FC}{Fog Computing}
\nomenclature{SNR}{Signal-to-Noise Ratio}
\nomenclature{V2X}{Vehicle-to-Everything}
\nomenclature{V2I}{Vehicle-to-Infrastructure}
\nomenclature{V2V}{Vehicle-to-Vehicle}
\nomenclature{DSRC}{Dedicated Short-Range Communication}
\nomenclature{RTT}{Round-Trip Time}
\nomenclature{PCA}{Principal Component Analysis}
\nomenclature{OOP}{Object-oriented Programming}
\nomenclature{UAV}{Unmanned Aerial Vehicle}
\nomenclature{HTTP}{Hypertext Transfer Protocol}

\begin{tcolorbox}[colframe=black,colback=white]
    \begin{multicols}{2} 
        \printnomenclature
    \end{multicols}
\end{tcolorbox}

\section{Introduction}

The Internet of Vehicles (IoV) has experienced remarkable growth and development in recent years, driven by the increasing demands of connected vehicles and the proliferation of for Intelligent Transportation Systems (ITS). According to a report by Allied Market Research, the IoV market is estimated to surge by 215\% by 2024 \cite{Pankaj2018}. In contrast to the Vehicular Cloud Computing (VCC), which is a centralized approach, VEC networks consist of distributed vehicles and roadside units (RSUs) all equipped with computing and storage resources. Within this context, VEC is emerging as a paradigm that enables the seamless integration of resources at the edge of vehicular networks \cite{10148943}. This opened new frontiers for real-time data processing and intelligent services in the vehicular domain, facilitating applications such as autonomous driving \cite{9580706}, security-enhanced Vehicular Ad Hoc Networks (VANETs) \cite{ARUL2023108905}, in-vehicle entertainment\cite{10.1145/3401979}, etc.

Resource scheduling trade-offs, particularly concerning service caching and offloading within its highly complex and dynamic network environment, pose the most pivotal challenge in VEC. In response to these challenges, edge intelligence scheduling \cite{9052677}, which refers to the intelligent paradigms, techniques, and problem solving of end device resource scheduling scheme, emerges in synergizing edge with Artificial Intelligence (AI) policies. These scheduling policies leverage Machine Learning (ML) to optimize Quality of Service (QoS) by taking into account factors such as the workload of edge devices \cite{8493149}, network congestion conditions \cite{GUO201940}, and energy consumption \cite{9960947}.

While attempts have been made to develop various VEC resource scheduling policies \cite{9832009,10034418}, researchers and industry practitioners encounter significant challenges when trying to evaluate them. Relying solely on real-world experimentation presents difficulties in terms of cost and time. Additionally, the inevitable interference of external environmental factors often compromises reproducibility, highlighting the significance of computer simulation technology. According to \cite{ROS20141}, simulators are software-based environments that can replicate the behavior of real-world systems, allowing for controlled experimentation, analysis, and evaluation of complex scenarios. Remarkably, a multitude of simulators have been developed across various domains, including server cache management \cite{10.1145/3372393}, Internet of Things (IoT) , Mobile Edge Computing (MEC) and fog networks.

As of the final draft of this paper, there is a notable absence of dedicated VEC simulators. VEC environments are distinguished by their substantial service deployment costs, emphasis on short-range communications, dynamically evolving network topologies due to vehicular mobility, stringent latency requirements, and predictable mobility patterns dictated by road infrastructure \cite{10.1145/3485129}. While some researchers have attempted to adapt existing simulators originally designed for MEC and Vehicle-to-Everything (V2X) network environments \cite{s22124580}, these approaches often fall short in fully capturing the distinct characteristics of VEC environments and potentially lead to inaccurate results. Specifically, OMNeT++ \cite{10.4108/ICST.SIMUTOOLS2008.3027} and NS-3 \cite{Riley2010} are both general-purpose network simulators that can be adopted to model a wide range of networks, including vehicular networks. However, these simulators are originally intended for broader scenarios and hampered by their heavy emphasis on compatibility, making them bulky and featuring a steep learning curve that hinders future customization.

In this paper, we propose VEC-Sim, a simulator meticulously designed for VEC networks. It provides researchers with a plug-and-play platform to seamlessly integrate their custom strategies and evaluate their performance within the simulated environment. VEC-Sim incorporates built-in support for all the key elements of VEC networks, including vehicular mobility, Vehicle-to-Infrastructure (V2I) communication \cite{s21030706}, and edge computing capabilities. Additionally, VEC-Sim features mechanisms to capture the unique characteristics of VEC networks such as latency sensitive, high mobility, distance-constraint and limited resources. VEC-Sim is also meticulously crafted to be user-friendly, featuring straightforward and intuitive APIs out of the box. This makes it accessible to a wide range of users, including researchers, developers, and students. Additionally, VEC-Sim is flexible and extensible, allowing users to easily customize it to meet their specific needs. 

The main contributions of this paper are summarized as follows:

\vspace{-\topsep}
\begin{itemize}
\setlength{\itemsep}{0pt}
\setlength{\parsep}{0pt}
\setlength{\parskip}{0pt}
    \item We conducted an analysis of vehicular network topologies and application characteristics. Models for communication, computation, and caching were developed to serve as a solid groundwork for subsequent simulator design.
    \item We designed a modular and layered architecture for VEC-Sim. Core entities like vehicles, services and connections were encapsulated in classes with well-defined interfaces. Rich APIs were provided to facilitate extension and customization by operators.
    \item We innovatively engineered an all-in-one solution capable of modeling caching and offloading tasks concurrently, which is crucial for evaluating joint resource scheduling policies in VEC networks. Besides, we constructed VEC-Sim based on time-slicing mechanism, ensuring both accuracy and ease of maintenance.
    \item To enhance the realism of the simulation, we proposed mechanisms such as spatial distribution generation, service interest vector, DRL-based vehicle mobility modeling, user preference shift and new service upload. Collectively, these mechanisms empower operators to replicate various complex factors inherent in vehicular network environments.
    \item In-depth experiments were conducted to showcase the capabilities of VEC-Sim, validating its feasibility and effectiveness in modeling VEC systems and performing case studies.
\end{itemize}
\vspace{-\topsep}

The reminder of the paper is organized as follows: Section 2 provides a brief overview of state-of-the-art simulators related to our work. Section 3 formulates the network model and Section 4 presents VEC-Sim's design details. Section 5 presents simulation experiments and case studies using VEC-Sim. Finally, Section 6 draws a conclusion of this paper and discusses the future research direction.

\section{Related Work}
In the field of computer network, the research community has made great efforts towards the development of simulation tools for testing and validation. With the ever-growing number of existing simulators, these tools exhibit variations in their intended usage, domain-specific features, functionalities, network infrastructure models, and even their open-source licensing agreements. Table \ref{tb1} presents a comparison of key characteristics between state-of-the-art simulators across various network domains.

\begin{table}[width=1.0\linewidth,cols=7,pos=h]
\caption{Comparison of VEC-Sim and well-established simulators in its relevant domains}\label{tb1}
\begin{tabular*}{\linewidth}{@{}p{0.14\linewidth}p{0.07\linewidth}p{0.15\linewidth}p{0.28\linewidth}p{0.07\linewidth}p{0.04\linewidth}p{0.1\linewidth}@{}}
\toprule
\textbf{Simulators} & \textbf{Language} & \textbf{Scope} & \textbf{Functionality} & \textbf{Mobility} & \textbf{GUI} & \textbf{License} \\ 
\midrule
OMNeT++ \cite{10.4108/ICST.SIMUTOOLS2008.3027} & C++ & Network stack & communication protocols,
  network topologies, traffic & Yes & No & GPLv2 \\
CloudSim \cite{10.1002/spe.995} & Java & Cloud computing & resource management, data
  center management, virtual machine management, load balancing & No & Yes & LGPLv2.1 \\
iFogSim \cite{Gupta2016iFogSimAT} & Java & Fog network & energy consumption,
  latency, response time, cost & Yes & Yes & Apache 2.0 \\
SimEdgeIntel \cite{WANG2021102016} & Java & Edge computing & device-to-device (D2D) communication, network switching mechanism, load balancing & Yes & Yes & LGPLv3 \\
IoTSim \cite{ZENG201793} & Java & IoT & IoT big data application,
 big data processing, latency and cost & Yes & Yes & MIT \\
 GrooveNet \cite{4141794} & C++ & Vehicular network & protocol design, in-vehicle deployment & Yes & Yes & GPLv2 \\
 \textbf{Ours} & Python & VEC & Benchmark of service caching and offloading policies &  Yes & Yes & MIT \\
\bottomrule
\end{tabular*}
\end{table}

One common approach in VEC domain research involves secondary development based on general-purpose simulator frameworks. For instance, OMNeT++ \cite{10.4108/ICST.SIMUTOOLS2008.3027} is frequently selected for customization because it comes pre-packed with a wide range of simulation models spanning diverse domains, including peer-to-peer networks, mesh networks \cite{PAULONJV2022102809}, wireless sensor networks, Storage Area Networks (SANs), and beyond. However, this approach often requires significant effort to adapt these general-purpose tools to the specific requirements of VEC networks, potentially leading to inefficiencies and reduced accuracy in modeling VEC-specific phenomena.

In terms of domain-specific simulators, simulators more closely aligned with our approach include those focused on Edge Computing (EC), Fog Computing (FC) and vehicular networks. To enable modeling of device mobility and service migration in fog computing environments, Puliafito et al. developed MobFogSim \cite{PULIAFITO2020102062}, an extension of iFogSim that facilitates evaluation of user movement patterns on application container migration. Shaik et al. proposed PFogSim \cite{SHAIK2022100736} to facilitate testing of hierarchical and geographically distributed fog infrastructures with varying fog node capacities. Nonetheless, PFogSim features coarse-grained modeling of edge application lifecycles. Jha et al. introduced IoTSim-Edge \cite{2020IoTSim}, extending CloudSim to simulate heterogeneous IoT and edge systems incorporating protocols, node mobility and resource management. Souza et al. presented EdgeSimPy \cite{SOUZA2023446} to accurately represent the entire lifecycle of edge applications from deployment to completion. In the realm of vehicular network simulation, a diverse array of tools has emerged to tackle the intricacies of modeling vehicular communication protocols and traffic dynamics. Mangharam et al. developed GrooveNet \cite{4141794}, a hybrid simulator that bridges virtual and real-world scenarios by enabling interactions among simulated and actual vehicles while incorporating diverse communication models within real street map topographies. Amoozadeh et al. created VENTOS \cite{AMOOZADEH201961}, an integrated VANET simulator for exploring traffic flow and collaborative driving, which implements features such as car-following models and platoon management protocols. Sommer et al. built Veins \cite{Sommer2019} based on OMNeT++ and SUMO, providing a bidirectionally coupled environment for concurrent network and mobility simulations, thus offering researchers a platform to develop and evaluate novel vehicular network protocols and applications.

While these simulators offer valuable insights into their respective domains, they exhibit significant limitations when applied to VEC environments. Existing EC and FC simulators \cite{Gupta2016iFogSimAT,WANG2021102016,PULIAFITO2020102062,SHAIK2022100736} typically lack comprehensive modeling of VEC ecosystems. These tools often struggle to capture the complex and rapid movements of vehicles, the dynamic nature of V2X communications, and the unique characteristics of RSU deployments. Additionally, these simulators generally fall short in representing the intricate interactions (e.g., service offloading, content caching \cite{10034418}, and resource allocation) between vehicles, RSUs, and cloud infrastructure that are fundamental to VEC scenarios. On the other hand, general-purpose vehicular network simulators \cite{4141794,AMOOZADEH201961,Sommer2019}, while proficient in modeling vehicle movements, lack advanced edge computing resource management functionalities. These simulators often fail to incorporate VEC-specific protocols and mechanisms (e.g., service migration \cite{9964436}, fault tolerance \cite{10014015}, caching), and encounter challenges in integrating and evaluating pertinent VEC-related performance metrics (e.g., end-to-end latency, resource utilization efficiency, QoE \cite{9422164}). Consequently, both categories of simulators struggle to accurately model the intricate interactions between vehicular mobility and edge resource allocation that are fundamental to VEC environments. These limitations underscore the necessity for a simulation platform that is meticulously designed for VEC systems.

\section{Network Model Formulation}
\subsection{VEC-empowered Network Topology}

\begin{figure}[!h]
        \centering
	\includegraphics[width=.95\linewidth]{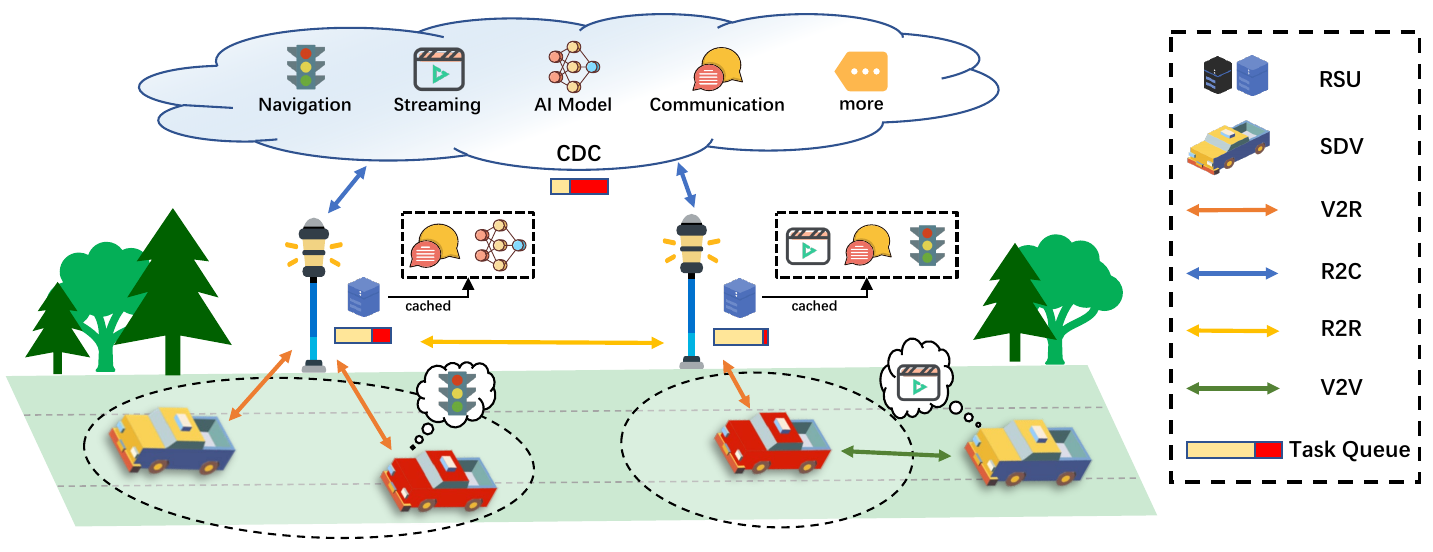}
	\caption{Network model of the VEC environment}
	\label{FIG:1}
\end{figure}

The proposed network model consists of three primary components: Cloud Datacenter (CDC), Roadside Units (RSUs), and Software-defined Vehicles (SDVs), as depicted in Figure \ref{FIG:1}.

The CDC serves as a centralized cloud platform hosting a repository of $K$ vehicular services denoted as $S = \{s_1, s_2, \ldots, s_K\}$. This collection of services encompasses a diverse range of application workloads, such as self-driving, over-the-air firmware updates, streaming media, etc. The VEC network facilitates elastic deployment and on-demand provisioning of each service, with the CDC providing high-performance computing resources to enable offloading of services for remote execution. The set of $H$ SDVs is represented as $V = \{v_1,v_2,\ldots,v_H\}$, where each $v_i$ has limited storage space and computational resources. SDVs utilize On-Board Units (OBUs) to request vehicular services from neighboring RSUs during travel via Dedicated Short-Range Communication (DSRC) technology. A discrete time frame structure is adopted such that $\tau \in \{1,2,\ldots,T\}$ indexes the simulation timeline. Thus, Service request issued by vehicle $v_i$ at frame $\tau$ for service $s_j$ can be denoted as ${req}_\tau^{v_i, s_j}$. RSUs are strategically positioned along roadways to facilitate data exchange between SDVs and the CDC. The set of $P$ RSUs is represented as $R = \{r_1,r_2,...,r_P\}$, where $r_i$ denotes the $i$-th RSU. Each RSU is assumed to have an associated edge server (ES) that acts as an edge computing node, providing localized storage and processing capabilities to RSU functions.   

Services occupy only storage space when not deployed. Upon receiving a service request from an SDV, the CDC executes an offloading policy that is externally integrated for evaluation by researchers. The offloading decision, based on status information such as network conditions, service requirements, and device constraints, can lead to three outcomes: cloud deployment (onto the CDC), edge deployment (onto an available RSU), or local deployment (directly onto the SDV). When deployed, services begin consuming CPU, RAM, and storage resources. To optimize resource utilization and reduce latency, researchers can optionally implement a caching strategy into VEC-Sim, where frequently requested service images can be stored at RSUs. This allows for faster deployment without the need to fetch the image from the CDC each time.

\subsection{Communication Model}
To enable flexible simulation of various network topologies, the communication model is constructed to include four types of connections: V2R (SDV to RSU), V2V (SDV2SDV), R2C (RSU to CDC) and R2R (RSU to RSU).

V2R and V2V communication are Point to Point (P2P) connections in a full-duplex manner \cite{s21030706}. Owing to the constrained wireless transmission power of RSUs and SDVs, it is essential that the requesting entity is within the effective communication range of its target. In a 2-D map, the coordinate of entity $(\cdot)$ can be represented as $l_x^{(\cdot)}$ and $l_y^{(\cdot)}$. The Euclidean equation is commonly used to calculate the distance, which can be expressed as

\begin{equation}
    \label{eq1}
    {dist}_{v,r}(\tau) = \sqrt{\left|lx_v(\tau) - lx_r(\tau)\right|^2 + \left|ly_v(\tau) - ly_r(\tau)\right|^2} .
\end{equation}

For RSUs with varying transmission powers, the coverage range may differ. Here, $g_i$ represents the coverage radius, with $i$ identifying a specific RSU. Then, the collection of entities capable of establishing communication with $r_i$ is denoted as $\mathbb{Z}^{r_i} \in \{z|{dist}_{z,r_i} \textless g_i\}$. Following Shannon's theorem \cite{1455040}, the maximum rate of clean data that can be transmitted over a channel is a function of the Signal-to-Noise Ratio (SNR). As many reflective surfaces present in urban environments, Rayleigh fading is then employed to model signal attenuation between two nodes \cite{601747}. To incorporate Rayleigh fading into the data transmission rate formula, we can modify the SNR to account for the additional fading effect. Then, the data transmission rate for the established connection $C^{a \leftrightarrow b}$ can be computed as

\begin{equation}
    \label{eq2}
    {tr}_{C^{a \leftrightarrow b}}(\tau) = \mathbb{B} \cdot \log_2\left(1 + \frac{{\min(\mathbb{P}_a, \mathbb{P}_b) \cdot \sigma \cdot {dist}_{a,b}(\tau) \cdot \mathbb{h}}}{{\mathbb{N}_0}}\right) ,
\end{equation}
where $\mathbb{B}$ is the channel bandwidth, $\mathbb{P}_{(\cdot)}$ denotes the transmission power, factor $\sigma$ is the path loss exponent, $\mathbb{h}$ is the model of the Rayleigh fading channel, and $\mathbb{N}_0$ is the amplitude of the Gaussian background noise.

R2C communication is utilized when there is a necessity for interactions with the CDC via backhaul connections. It is pertinent to acknowledge that the outputs of service execution typically exhibit notably smaller data volume in contrast to input parameters \cite{9284036}. Thus, we assume the response transmission delay tends to be inconsequential. The primary time consumption arises from the propagation delay resulting from the relatively long geographical distance between the CDC and RSUs, which is independent of the amount of data being transmitted. In this context, the time consumption can be approximated using Round-Trip Time (RTT) as

\begin{equation}
    \label{eq3}
    {RTT} \approx 2 \cdot t_{R2C}^{prop} = \frac{{dist}_{r,CDC}}{{v_{tran} \cdot \zeta}} ,
\end{equation}
where $v_{tran}$ is the velocity of electromagnetic signal propagation, and $\zeta$ represents the reduction in propagation speed due to signal attenuation within the transmission medium.

VEC-Sim also incorporates multi-hop support with R2R mesh communication, thereby allowing for in-depth analysis of node collaboration strategies \cite{9047880} and comprehensive testing of fault tolerance mechanisms. For unreachable RSUs $r_a \notin \mathbb{Z}^{r_b}$, mesh networking $\mathbb{M} = \{r_a \leftrightarrow r_n, \ldots, r_m \leftrightarrow r_b\}$ provides an alternative means of connectivity through relaying. The time consumption to complete $C^{r_a \leftrightarrow r_b}$ through $\mathbb{M}$ can be calculated as

\begin{equation}
 T^{mesh}_{r_a \leftrightarrow r_b}(\tau) = \sum_{\otimes \in \mathbb{M}} \left[ \frac{{size}(C^{\otimes})}{{tr}_{C^{\otimes}}(\tau)} + t_{\otimes}^{prop} \right] .
\end{equation}

In standard operational scenarios, SDVs establish communication with the CDC through a two-hop process. Initially, they connect to a proximate RSU via V2R communication. Subsequently, the RSU relays the data to the CDC through its backhaul link, employing R2C communication. In the event of RSU backhaul link failure to the CDC, R2R communication is utilized to establish ad-hoc routing paths. This approach effectively mitigates the impact of the failure point, thereby ensuring system connectivity.

\subsection{Computation Model}

Due to the high mobility of VEC entities, choosing the optimal back-end for executing computational tasks while adhering high Quality of Service (QoS) poses a significant challenge. We assume that $v_n$ issues an offloading task, described as ${task}_{v_n}^\tau = \langle \delta, \eta, \Delta t_{max} \rangle$, where $\delta$ is the required CPU cycles for processing, $\eta$ represents the service size, and $\Delta t_{max}$ denotes the timeout duration. For a given task, each SDV's service request can only be offloaded and executed at one of the following locations: locally within the SDV, at an RSU, or within the CDC.

\begin{enumerate}[(a)]
    \item \textbf{Local Offloading:} This pertains to the computational capacity within the SDV itself, designed to process tasks without external offloading. This approach is particularly advantageous when the network conditions are poor or the computational task is not resource-intensive. The available local computational capacity of $v_n$ at time $\tau$ is denoted by $f_{v_n}(\tau)$, measured in Floating-Point Operations Per Second (FLOPS). Then, the time required for local processing can be calculated as
    \begin{equation}
        T_{SDV}(\tau) = f_{v_n}(\tau) .
    \end{equation}

    \item \textbf{RSU Offloading:} This involves harnessing the computational capabilities of the units located at the RSU, which reduces the load on the SDVs and provides lower latency compared to CDC offloading due to the proximity of ES. However, each RSU can only concurrently handle a finite number $\omega$ of tasks \cite{9284036}. To prevent the system from getting overwhelmed, excessive requests are queued as $Q_\tau = \{{task}_1, {task}_2, \ldots\}$. The queuing time at $r_{{target}}$ can be expressed as
    \begin{equation}
        t_{r_{target}}^{queue}(\tau) =
        \begin{cases}
        0,& curr\_reqs<\omega \\
        \sum_{n=1}^{{len}(Q_\tau)} {duration}({task}_n),& {curr\_reqs} \geq \omega
        \end{cases} . 
    \end{equation}
    Therefore, the total time consumption of offloading to an RSU includes network transmission, queuing, and processing time. The formula for calculating the overall time consumption is
    \begin{equation}
        T_{RSU}(\tau, \delta) = t_{v_n \leftrightarrow r_{target}}^{trans}(\tau) + t_{r_{target}}^{queue}(\tau) + \frac{\delta}{f_{r_{target}}(\tau)} .
    \end{equation}

    \item \textbf{CDC Offloading:} This is typically used for offloading computationally demanding tasks that exceed the capabilities of local SDVs or RSUs. In this process, the parameters are first uploaded from $v_n$ to $r_{target}$, then forwarded via $r_{target}$ to the CDC for processing. The transmission of response follows the reverse path, going from CDC back through $r_{target}$ to $v_n$. Thus, the feedback time can be calculated as
    \begin{equation}
        T_{CDC}(\tau, \delta) = t_{v_n \leftrightarrow {CDC}}^{trans}(\tau) + \frac{\delta}{\Phi_{{task}_{v_n}^\tau} * f_{CDC}(\tau)} + {RTT} ,
    \end{equation}
    where $\Phi_{{task}_{v_n}^\tau}$ is the number of computational resources allocated to the task, and $f_{CDC}(\tau)$ is the computational capability of the CDC.
\end{enumerate}

To represent service offloading destinations, Boolean variables $\alpha , \beta \in \{0, 1\}$ are employed, where $\alpha$ and $\beta$ are defined as the decision variables indicating whether to offload locally or to the RSU, respectively. The overall computation time can be calculated as

\begin{equation}
    T_{{task}_{v_n}^\tau} = \alpha * T_{SDV} + (1-\alpha)\left[\beta * T_{RSU} + (1-\beta)*T_{CDC}\right] .
\end{equation}

\subsection{Caching Model}\label{section_3.4}

\begin{figure}[!h]
        \centering
        \includegraphics[width=.65\linewidth]{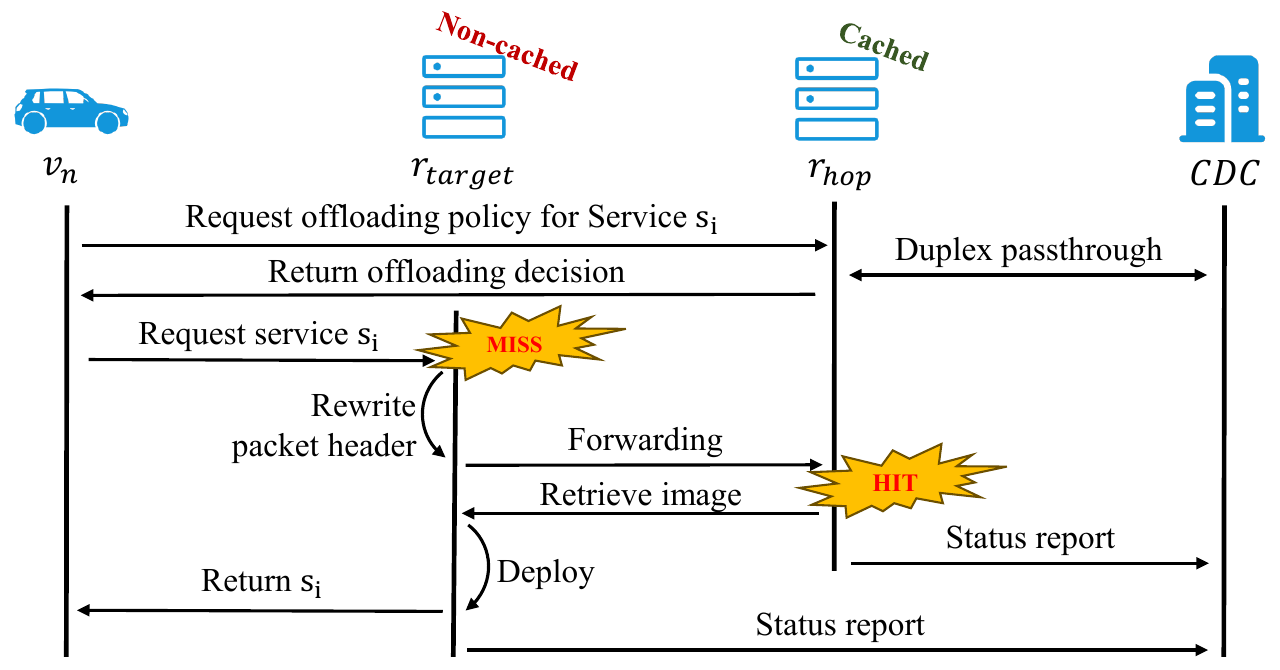}
	\caption{Request sequence diagram illustrating the cases of cache miss and cache hit, where the user has customized an inter-RSU collaboration mechanism. If the selected RSU lacks the required service image in its cache, it retrieves the image from nearby RSUs with cached copies.}
	\label{FIG:2}
\end{figure}

RSUs make decisions on whether to cache requested content in their local storage based on predetermined policies. This strategy enables RSUs to directly retrieve content from cache when other SDVs request the same service, thereby reducing data transmission latency and network load. Figure \ref{FIG:2} illustrates a typical timing sequence of $v_n$ requesting service $s_i$, demonstrating both cache miss and hit scenarios.

Specifically, $v_n$ initially sends a request for service $s_i$’s offloading policy to the CDC through $r_{hop}$, the RSU with the highest link quality. The CDC processes this request and returns the result, which includes the offloading target $r_{target}$ and nearby RSUs with cached content. This response is then relayed back to $v_n$ through $r_{hop}$. Upon receiving the offloading instructions, $v_n$ attempts to deploy $s_i$ on $r_{target}$. If $r_{target}$ does not have the required image cached locally (a cache miss), it initiates the user-defined inter-RSU collaboration mechanism. $r_{target}$ forwards the request to $r_{hop}$, which has the image cached, resulting in a cache hit. In scenarios where no RSU within range has the required image cached, the system retrieves it from the CDC. Finally, $s_i$ is deployed on $r_{target}$ and the service result is returned to $v_n$.

As observed in Figure \ref{FIG:2}, the system encounters both cached and non-cached scenarios, predominantly affecting the network transmission time. Let $\gamma_i$ denote whether the required content $s_i$ is cached. Since the CDC uses backhaul links, the transmission delay between CDC and RSUs is negligible. Thus, the time consumption on service request between $v_n$ and CDC can be expressed as

\begin{equation}
   T_{v_n \leftrightarrow {CDC}}^{req} \approx T_{v_n \leftrightarrow r_{{target}}}^{req} = RTT + (1 - \gamma_i) \left[ \frac{{size}(s_i.{image})}{tr_{C_{r_{{target}} \leftrightarrow r_{hop}}}(\tau)} + t_{r_{{target}} \leftrightarrow r_{{hop}}} ^{prop} \right] + T_{deploy} .
\end{equation}

\section{The Design of VEC-Sim}

\begin{figure}[!h]
	\centering
		\includegraphics[width=0.7\linewidth]{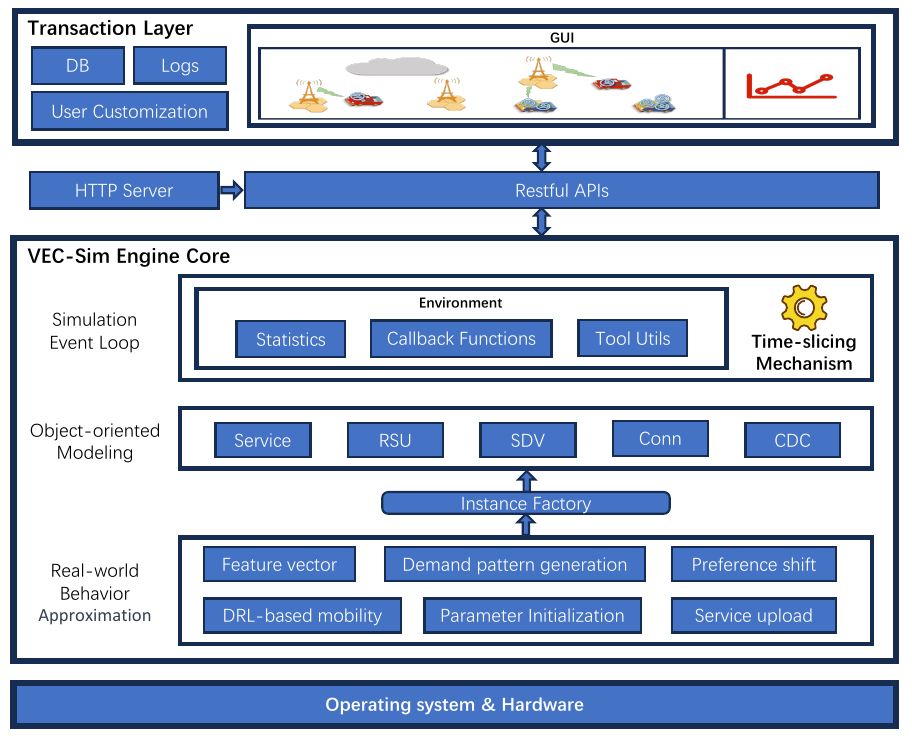}
	\caption{Architecture of VEC-Sim}
	\label{FIG:3}
\end{figure}

VEC-Sim is a simulation platform specifically developed to address resource scheduling, service caching, and task offloading challenges within the context of VEC environments. Besides, VEC-Sim offers the capability to replicate real-world network scenarios, encompassing scenarios such as cold starts, high concurrency, server downtime, and stress testing. The layered architectural design of VEC-Sim ensures modularity and scalability, providing a framework for seamless expansion and customization to address specific research requirements. Figure \ref{FIG:3} provides an overview of the VEC-Sim architecture, elucidating the intricate relationships between various components and their interactions. 

In this section, we first explain our choice of development language and establish the fundamentals by defining variables and initialization procedures. We then delve into the specific algorithms and models implemented within VEC-Sim, including the synthetic demand generation, vehicle mobility modeling, etc. Finally, we delve into the time-slice mechanism backbone which provides a controlled and elegant design for advancing the simulation in a flexible, decoupled and hardware-independent manner.

\subsection{Definition of Classes, Attributes, and Methods}

Python was selected as VEC-Sim's primary language for its versatility, readability, and rich ecosystem. Python's syntax simplicity and dynamic typing facilitate rapid prototyping, making it ideal for VEC-Sim users to integrate their own caching and offloading algorithms. This flexibility is further enhanced by Python's rich ecosystem of libraries, which provides essential tools for various aspects of the project, including scientific computing with NumPy, machine learning with scikit-learn, TensorFlow, and PyTorch, and data visualization using Matplotlib. Besides, its widespread adoption in the scientific community has led to an extensive collection of well-established baseline algorithms \cite{XU2023102780} implemented using Python. This prevalence allows for seamless comparisons with existing work and facilitates the integration of state-of-the-art techniques into VEC-Sim.

While Python may introduce some performance overhead compared to compiled languages like Java or C/C++, it offers solutions to mitigate this drawback. Python's ability to integrate with C/C++ allows for optimization of performance-critical sections when necessary. Furthermore, Python's support for concurrent programming paradigms such as processes, threads, and coroutines provides additional avenues for performance enhancement, making it a well-rounded choice for VEC-Sim's development.

\begin{figure}[!h]
	\centering
		\includegraphics[width=1\linewidth]{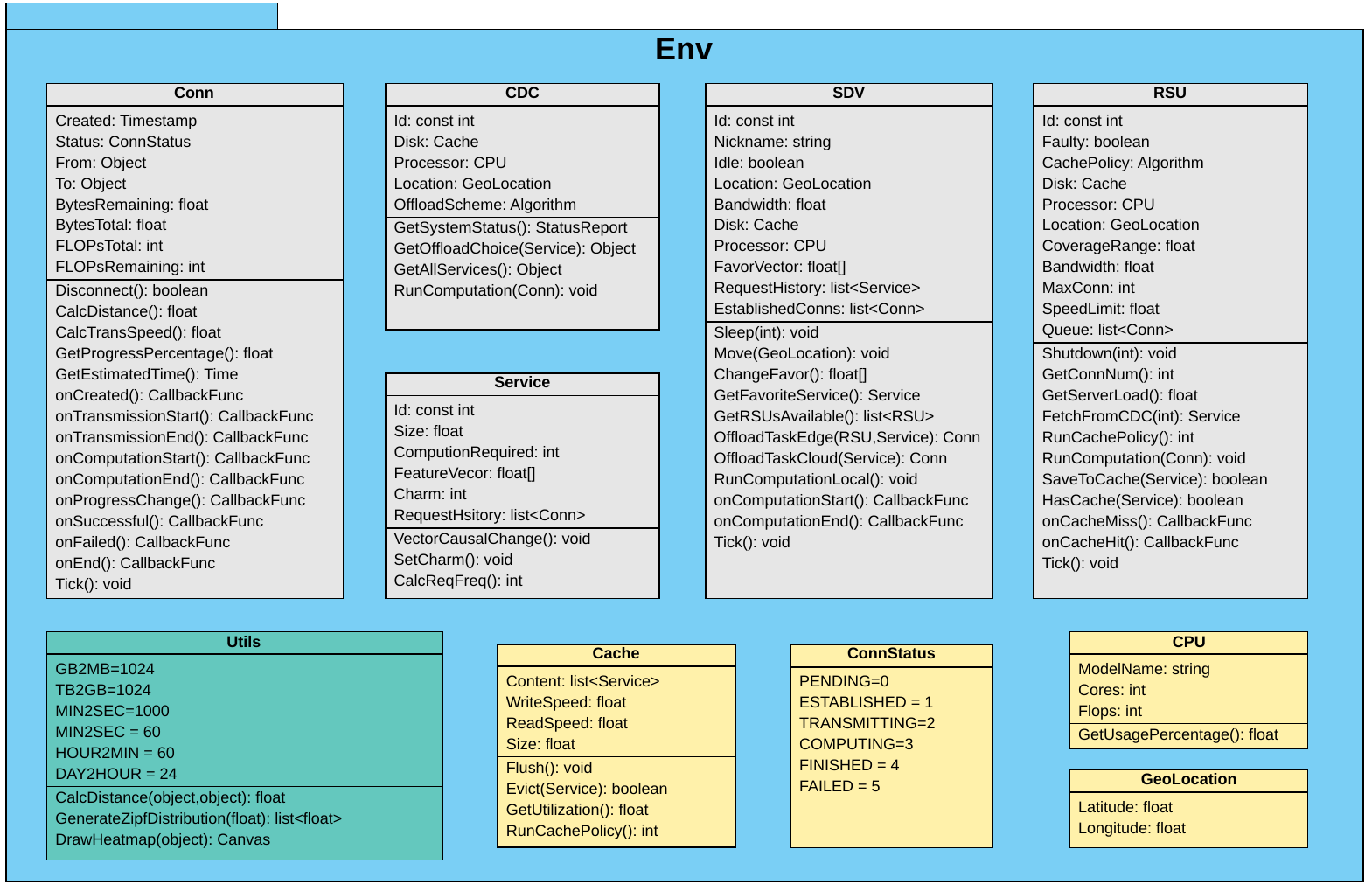}
	\caption{Class diagram of VEC-Sim’s $Env$ package, where the gray boxes (e.g., $Conn$, $CDC$, $SDV$, etc.) represent the primary classes that model the core entities and functionalities of the VEC environment. The yellow boxes (e.g., $Cache$, $ConnStatus$) represent auxiliary classes or data structures that are used by or related to the primary classes.}
	\label{FIG:4}
\end{figure}

VEC-Sim is implemented using Object-oriented Programming (OOP) techniques, which organizes state and behaviors into modular, reusable classes. This enables us to replicate the complexity of real-world VEC systems within VEC-Sim. Researchers can also leverage this well-defined architecture to customize simulation entities and extend self-defined functionality with custom behaviors tailoring to their specific research objectives. Specifically, this flexibility allows for the implementation of user-defined scheduling policies \cite{9580706,9960947}, V2X service access protocols (Figure \ref{FIG:2}), and the creation of custom traffic generation models to simulate specific scenarios (e.g., rush hour congestion, UAV-assisted communication \cite{10077418}). The attributes and methods of various entities within the modular VEC-Sim architecture are represented using Utilizing Unified Modeling Language (UML) diagram, as depicted in Figure \ref{FIG:4}.

The package labeled $Env$ serves as a comprehensive context of the overarching simulation environment, where all real-world entities and their interactions occur. Within $Env$, we encounter the $SDV$, $RSU$, $CDC$, $Service$ and $Conn$ objects, designed to mirror their real-world counterparts. Details of these classes are present as follows.

\vspace{-\topsep}
\begin{itemize}
\setlength{\itemsep}{0pt}
\setlength{\parsep}{0pt}
\setlength{\parskip}{0pt}
    \item $\mathbf{SDV}$: Instances of the $SDV$ class are entities that feature mobility, can request services, and possess a certain amount of computational and storage resources.
    \item $\mathbf{RSU}$: Instances of the $RSU$ class are fixed infrastructures that aid in communication, caching, and computation. They provide interfaces such as $GetConnNum()$ and $GetServerLoad()$ to obtain resource utilization information, which serves as input for evaluating strategies. The $CachePolicy$ variable allows the integration of custom service caching strategies.
    \item $\mathbf{CDC}$: The $CDC$ class stores all services within the system in its member variable $Disk$. It is responsible for the service offloading decisions across the system. Users can replace its $OffloadScheme$ to implement their own scheduling strategies.
    \item $\mathbf{Service}$: This class represents services that can be cached and executed in the VEC environment, with attributes for service characteristics and computation requirements.
    \item $\mathbf{Conn}$: When data exchange occurs between these entities, a corresponding $Conn$ object is instantiated to replicate the network connection. Throughout its lifecycle, the $Conn$ object's status can be \texttt{PENDING}, \texttt{ESTABLISHED}, \texttt{TRANSMITTING}, \texttt{COMPUTING}, \texttt{FINISHED}, and \texttt{FAILED}. Notably, the $Conn$ object manages the lifecycle of a connection, emitting events related to data transmission and computation until the action is completed.
\end{itemize}
\vspace{-\topsep}

\begin{figure}[!h]
	\centering
		\includegraphics[width=1\linewidth]{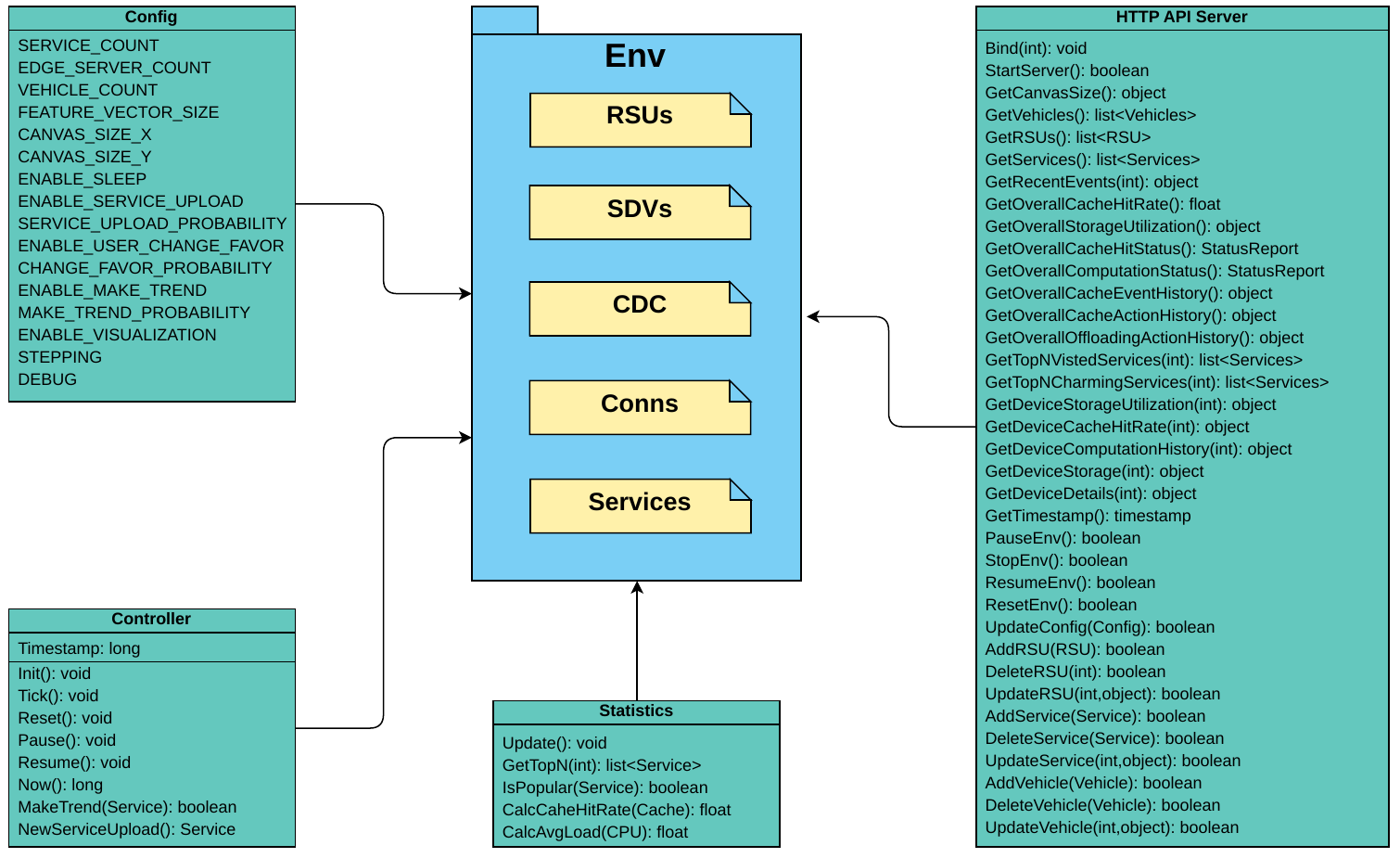}
	\caption{Supporting modules working alongside with the core $Env$ package.}
	\label{FIG:4_new}
\end{figure}

Beyond the core Env package, several supporting modules (Figure \ref{FIG:4_new}) are implemented to enhance the functionality of VEC-Sim to provide a user-friendly GUI, configurability, real-time control, and statistical reporting capabilities.

\vspace{-\topsep}
\begin{itemize}
\setlength{\itemsep}{0pt}
\setlength{\parsep}{0pt}
\setlength{\parskip}{0pt}
    \item $\mathbf{HTTP ~ API ~ Server}$: This interface extends the accessibility of VEC-Sim. This interface enables external systems and users to programmatically interact with VEC-Sim through Hypertext Transfer Protocol (HTTP) requests, facilitating remote control and comprehensive monitoring of the simulation process.
    \item $\mathbf{Controller}$: The $Controller$ object plays a pivotal role in managing the simulation progression, empowering researchers to re-initiate, pause, resume, make trends, and services upload to the simulation context.
    \item $\mathbf{Statistics}$: This module gathers and processes simulation data, providing insights such as top services, popularity metrics, cache hit rates, and average loads.
\end{itemize}
\vspace{-\topsep}

\subsection{Environment Parameter Initialization}
In the realm of simulation, proper initialization of parameters is paramount to replicate real-world scenarios accurately. Within this context, we focus on three primary areas of initialization:

\vspace{-\topsep}
\begin{enumerate}[(a)]
    \item \textbf{Service Size, Charm and Computational Load Distribution:} Aforementioned properties all exhibit the "long tail" phenomenon, where a small subset of elements possess significantly higher frequency or impact compared to the majority. Empirical evidence and prior research \cite{GHASEMI202211} confirm that service sizes have been observed to follow power-law distributions. According to Cisco’s Global IP Traffic Forecast \cite{cisco2022}, few highly popular services dominate approximately 80\% of internet traffic, while less-visited resources exhibit stable and persistent access amount that does not diminish to zero. In practice, this means that a few highly popular services account for the majority of service requests, while a multitude of niche services exist within the long tail of the distribution. These distribution patterns can be effectively modeled using the Zipf distribution, which has been validated in numerous prior works \cite{tian2022dima}. The Zipf distribution is mathematically defined using Probability Mass Function (PMF) as

    \begin{equation}
        P(X=k,\alpha)=\frac{1}{k^\alpha \cdot H_N} ; ~~ \text{with} ~ H_N = \sum_{n=1}^N \frac{1}{{n^\alpha}} .
    \end{equation}
    
    Here, $P(X=k,\alpha)$ represents the probability that the random variable $X$ takes on the value $k$, and $H_N$ denotes the $N$-th harmonic number. The parameter $\alpha$ is the Zipf factor, which determines the degree of skewness in the distribution. 
    

    \item \textbf{SDV Cluster and RSU Placement:} In real-world VEC systems, SDVs and RSUs are strategically distributed across various regions of the map. The coverage ranges of RSUs may overlap each other, forming hotspots in certain areas. SDVs can reside within the communication ranges of multiple RSUs. Moreover, due to the dynamic nature of vehicular environments and other factors, the density of SDVs surrounding RSUs can vary significantly. The Location Cluster Generation method of Algorithm \ref{alg1} allows researchers to independently specify cluster density constraints and entity amounts. By controlling the density of SDVs in each cluster, it can model the non-uniform densities that SDVs exhibit surrounding infrastructure in different parts of the network. Besides, the locations of SDVs and RSUs are generated stochastically to simulate the irregular coverage patterns that emerge from overlapping and sparse transmission zones of RSUs.

    ~\\
    \begin{algorithm}[H]
        \SetAlgoLined
        \SetKwInOut{Input}{Input}
        \SetKwInOut{Output}{Output}
        \Input{Canvas size $(X,Y)$, SDV count $m$, RSU count $n$, Cluster density set $D=\{d_1,d_2,\ldots,d_n\}$}
        \Output{Entity cluster $\{R,S\}$}
        \caption{Location Cluster Generation}
        \label{alg1}
        
        $R = \{r_1,r_2,\ldots,r_n\}$ \quad// {Initialize $R$ to target length}\\
        $S = \{\varnothing\}$ \quad// {Initialize $S$ as empty set}\\
        
        /* Randomly place RSUs */\\
        \For{$\textbf{each } r_i \textbf{ in } R$}{
            $r_i.{Latitude} = {random\_in\_range}(0,X)$\\
            $r_i.{Longitude} = {random\_in\_range}(0,Y)$\\
        }
        
        \For{$(i=1;i \leq n;i+=1)$}{
            /* Convert density to actual SDV count per cluster */\\
            $num = {round}(m \cdot \frac{d_i}{\sum_{k=1}^n d_k})$\\
            \For{$(j=1;j \leq num;j+=1)$}{
                Instantiate the corresponding SDV entity $s$\;
                /* Generate using polar coordinates */ \\
                $radius = {random\_in\_range}(0,r_i.{CoverageRange})$\\
                $angle = {random\_in\_range}(0,2\pi)$ \quad// {360 degrees}\\
                $s.{Latitude} = r_i.{Latitude} + radius \cdot \cos(angle)$\\
                $s.{Longitude} = r_i.{Longitude} + radius \cdot \sin(angle)$\\
                $S = S \cup \{s\}$\\
            }
        }
    \end{algorithm}
    ~\\

    \item \textbf{Service Feature Vector Generation:} To effectively simulate the service access pattern of real-world SDVs, we employ the service feature vector mechanism inspired by recommendation systems \cite{9269396}. Each service vector comprises 128 dimensions, serving as a high-dimensional encoding of the service characteristics. Notably, services of the same type exhibit higher similarity in their respective service vectors. Algorithm \ref{alg2} outlines the methodology for generating service feature vectors. Lines 4-12 begin by creating a central vector corresponding to each service type. Next, lines 17-22 employ a Gaussian distribution to generate each service feature, leveraging the inherent similarities among the same category. Finally, the algorithm generates the desired number of vectors according to user specification.

    ~\\
    \begin{algorithm}[H]
        \SetKwInOut{Input}{Input}
        \SetKwInOut{Output}{Output}
        
        \caption{Service Feature Vector Generation}
        \label{alg2}
        
        \Input{Service number $g$, Dispersion $m$, Cluster count $f$, Vector length $v$}
        \Output{Service vector set $S$}
        
        $S = \{s_1, \ldots, s_n\}$ \quad// {Initialize $S$ to the target length}\\
        $C = \{\emptyset\}$ \quad// {Define the center service vector set}\\
        $A = \{\emptyset\}$ \quad// {Define the number of services included in each service type}\\
        
        
        /* Generate center vectors for each service */ \\
        \For{$(i = 1; i \leq f; i += 1)$}{
            $X = [1, \ldots, v]$ \quad// {Define a temporary service vector}\\
            \For{$(k = 1; k \leq v; k += 1)$}{
                $X[k] = {random\_in\_range}(0, 10)$\\
                $c = X, a = \frac{g}{f}$\\
                $C = C \cup \{c\}$, $A = A \cup \{a\}$\\
            }
        }
        
        
        /* Generate service features according to each center */ \\
        \ForEach{$a_i, c_i$ in $[A, C]$}{
            \For{$(k = 1; k \leq a_i; k += 1)$}{
                $X = [1, \ldots, v]$\\
                \For{$(j = 1; j \leq v; j += 1)$}{
                    /* Each feature follows a Gaussian distribution */ \\
                    $p(x_j) = {random\_in\_range}(0.8, 1)$\\
                    $t = \sqrt{-2m^2 * \ln \left(\sqrt{2\pi m^2} p(x_j)\right)}$\\
                    $x[j] = {random\_in\_range}(c_i[j] - t, c_i[j] + t)$\\
                }
                $s_{k + \sum_{n=0}^{i-1} a_n}.\text{{vector}} = X$\\
                $s_{k + \sum_{n=0}^{i-1} a_n}.\text{{type}} = i$\\
            }
        }
        
    \end{algorithm}

\end{enumerate}
\vspace{-\topsep}

\subsection{Synthetic Demand Pattern Generation} \label{section_4.3}

To investigate the traffic demands of general contents, we utilized the traffic data collected at a Content Delivery Network (CDN) cluster of Akamai across North America \cite{8552665}. By analyzing the traffic demands patterns over a one-week period, our investigation unveiled several noteworthy findings: \textbf{a) Seasonal Traffic Patterns:} There was a noticeable surge in traffic during specific times of the day. This can be attributed to user internet usage habits. \textbf{b) User Preference-Driven Content Demands:} Content popularity was influenced by users' preferences and interests. \textbf{c) Impact of Special Events:} Certain special events, such as emerging hot topics, led to a substantial increase in traffic volume.

To simulate real-world SDVs making on-demand service requests, each SDV features a preference vector $h_v$ of dimensionality 128, and each service is characterized by a corresponding feature vector $f_s$ of the same dimensionality. When users are idle, they randomly select 50 services and traverse each service to calculate the cosine similarity between their preference vector $h_v$ and the service's feature vector $f_s$. By combining the service's charm value $a_s$, we can then obtain an interest score that reflects the user's preference for that service. Finally, the user will always select the service with the highest interest score to make a request. The process of calculating the interest score $I_{v,s}$ can be formulated as

\begin{equation}
    \label{eq11}
    I_{v,s} = a_s \cdot \cos(h_v, f_s) = a_s \cdot \frac{h_v \cdot f_s}{|h_v| \cdot |f_s|} .
\end{equation}

Algorithm \ref{alg3} outlines the process of service selection for SDVs. Specifically, Lines 2-9 replicate the behavior of users randomly opening a webpage and viewing $num$ services for selection. Line 10 combines the randomly selected services with the hot-ranking list to form a unified set of candidates. Lines 12-18 iterate through this set and calculate the interest score $I_{v,s}$. For services that have already been accessed by the SDV, a discount factor $\epsilon$ is applied to adjust their interest score accordingly. Finally, the service with the highest interest score is returned as the service to request.

    ~\\
    \begin{algorithm}[H]
        \SetKwInOut{Input}{Input}
        \SetKwInOut{Output}{Output}
        
        \caption{Service Selection}
        \label{alg3}
        
        \Input{Service set $S = \{s_1, s_2, \ldots, s_K\}$, Hot-ranking list $H$, SDV interest vector $h_v$, Window size $num$, Discount factor $\epsilon$}
        \Output{Service to request $s_{target}$}
        
        $D = \{\varnothing\}$ \quad// {Initialize an empty set to store services}
        
        
        $count = 0$\\
        \While{$(count < num)$}{
            Randomly select service $s$ from $S$\;
            \If{$s \notin D$}{
                $D = D \cup \{s\}$\\
                $count++$\\
            }
        }
        $D = D \cup H$ \quad// {Combine services from the hot-ranking list}
        
        
        $temp = \{\varnothing\}$\\
        \ForEach{service $s$ in $D$}{
            Calculate interest score $I_{v,s}$ using Eq. (\ref{eq11});\\
            \If{$s$ has not been previously accessed by SDV}{
                $I_{v,s} = I_{v,s} \cdot \epsilon$\\
            }
            Append $s$ to $temp$ while maintaining descending order of $I_{v,s}$\;
        }
        
        
        \textbf{Return} $\max(temp)$ as the service to request\\
        
    \end{algorithm}
    ~\\

VEC scenarios differ significantly from traditional desktop user behaviors as users temporarily disengage from further interactions due to the dynamic nature of vehicular environments \cite{10.1145/3485129}. To replicate this interaction dynamics, a spatial-temporal sleep mechanism is introduced where SDVs may enter an inactive state after issuing a service request. Let $\mathcal{U}(a,b)$ denote a uniform distribution between $a$ and $b$. The SDV’s sleep duration can be modeled as

\begin{equation}
T_{sleep} \sim k \cdot \mathcal{U}\left(e^{|\theta_a|}, e^{|\theta_a|} + \log_{1+\sigma}(\theta_v)\right) ,
\end{equation}

where $k$ is a scaling factor that adjusts the overall duration of the sleep period, and $\sigma$ reflects how speed impacts the range of the sleep duration. Additionally, $\theta_a$ and $\theta_v$ represent the current acceleration and velocity of the SDV, respectively.

This model captures the strong correlation between user interactions and vehicle dynamics. The incorporation of an exponential function ensures that even minor changes in acceleration translate into substantial adjustments in sleep duration. This design reflects the reality that periods of high acceleration or deceleration (indicated by a sharp increase in $|\theta_a|$) typically indicate active engagement in vehicle control, necessitating longer intervals between user interactions. Besides, the model expands the range of possible sleep duration as speed $\theta_v$ increases. This expansion accounts for the tendency of users to engage in passive activities (e.g., listening to podcasts) during high-speed travel. Such activities are characterized by extended periods of minimal interaction, punctuated by occasional engagement with the system.

In addition, several supplementary functionalities have been integrated to create more realistic simulations. We previously recognized that real-world users temporarily disengage from further interactions. To replicate this behavior, a sleep mode mechanism is introduced where SDVs may enter an inactive state after issuing a certain number of requests. SDVs' preferences can change over time due to various factors like marketing campaigns or evolving trends. Thus, we designed a mechanism to simulate these preference shifts by introducing accumulative Gaussian noise into $I_{v,s}$. Lastly, VEC-Sim also offers researchers the option to periodically add new services into the environment. This mirrors the continual expansion of service providers in the real world.

\subsection{Simulation of SDV Mobility Trajectories} \label{section_4.4}

\begin{figure}[!h]
	\centering
		\includegraphics[width=0.75\linewidth]{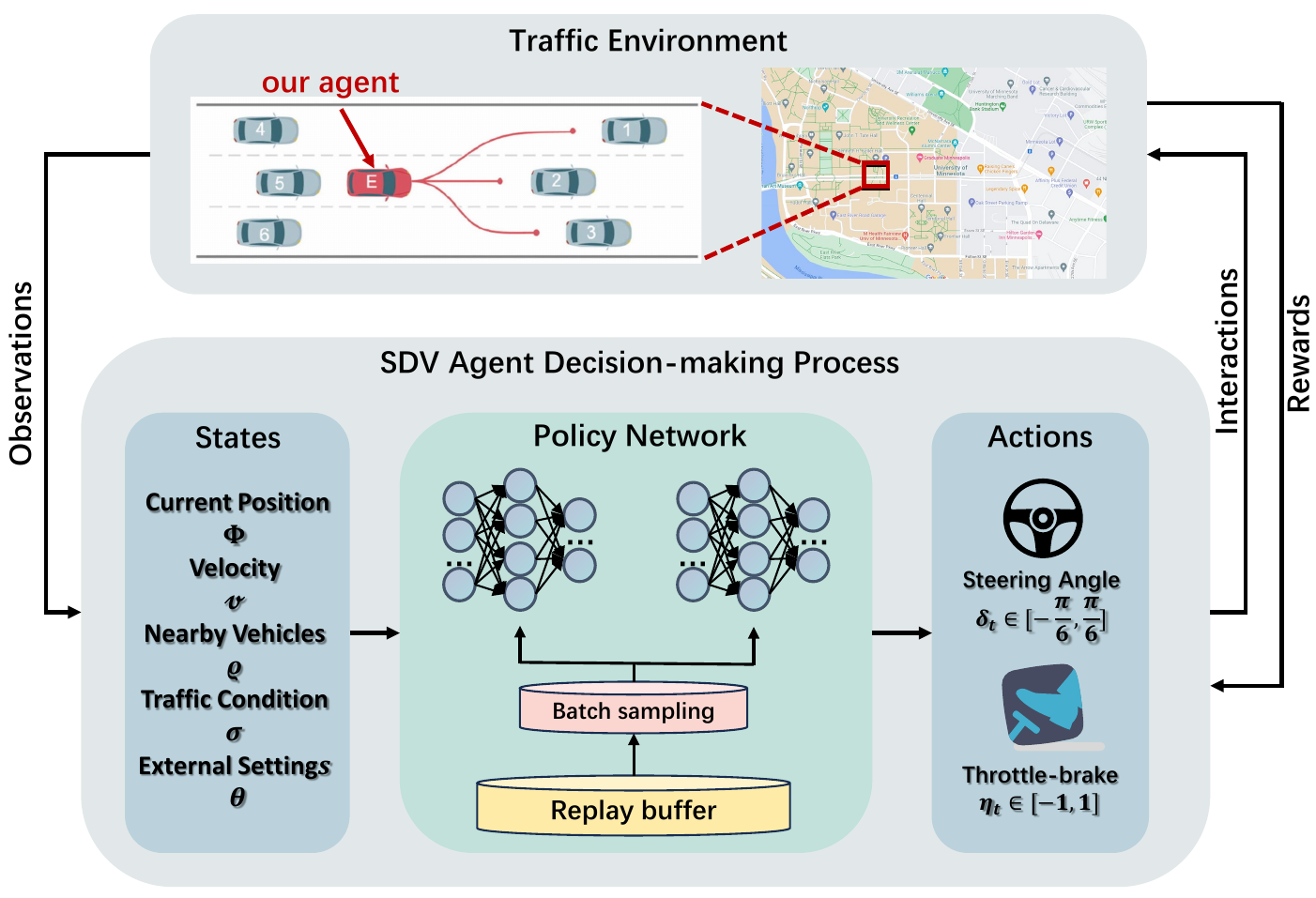}
	\caption{DRL-based SDV mobility simulation framework}
	\label{FIG:5}
\end{figure}

Inspired by the recent advancements of Deep Reinforcement Learning (DRL) in autonomous driving \cite{BADUE2021113816}, VEC-Sim adopts a DRL-based approach to model the mobility trajectories and driving behaviors of SDVs. Each SDV is modeled as a DRL agent that learns its optimal driving policy through interactions with the simulated environment. The high-level architecture of the proposed SDV mobility simulation framework is depicted in Figure \ref{FIG:5}.

Our initial step involves the creation of a virtual urban road environment. To achieve this, we directly employ Google Maps' API, which enables the retrieval of traffic data, including road layouts, real-time traffic congestion updates, and intersection regulations. Subsequently, the formalization of our agents' behavior hinges on the systematic definition of a Markov Decision Process (MDP) consisting of three fundamental components \cite{9885774}:

\vspace{-\topsep}
\begin{enumerate}[(a)]
    \item \textbf{State space:} The state space encapsulates the agent's observation of the environment at time $t$, which comprises: 1) The current position $\Phi$ represented using geographical coordinates in terms of longitude and latitude. 2) Velocity of the SDV $\mathscr{v}$ as a numerical value. 3) Nearby vehicles encoding $\varrho$, obtained by transforming relevant attributes of neighbors into a high-dimensional vector using a Graph Neural Network (GNN) encoder \cite{10.1145/2939672.2939754,9157331}. 4) Traffic condition indicator $\sigma$, where $\sigma = 0$ indicates free-flowing traffic, $\sigma = 1$ suggests moderate traffic congestion, $\sigma = 2$ signifies heavy traffic congestion and $\sigma = 3$ represents adverse conditions like accidents or road closures. 5) External motion settings $\theta$, which include user settings such as the target velocity and the frequency of lane changes. Hence, we denote the observation at time $t$ as $S_t = [\Phi, \mathscr{v}, \varrho, \sigma, \theta]$.

    \item \textbf{Action space:} At a given time step $t$, the policy network output is represented as $\mathcal{A}_t = \{\delta_t, \eta_t\}$, where $\delta_t \in [-\pi/6, \pi/6]$ encodes the steering angle in radians, and $\eta_t \in [-1, 1]$ denotes the linear throttle-brake signal. Notably, $\eta_t=1$ signifies full throttle, while $\eta_t=-1$ corresponds to full brake.

    \item \textbf{Reward function:} The reward function serves two primary purposes: 1) encouraging the vehicle to maintain a speed close to the target velocity and 2) adhere to the target frequency of lane changes for achieving customizable SDV motion parameters. Additionally, to make the simulation more realistic, the reward function also considers other important driving factors such as staying in the lane, smoothness of movement, and reasonability of actions accounting for traffic conditions and potential collisions. These factors are combined into a comprehensive reward signal to shape the desired SDV driving behavior.

\end{enumerate}
\vspace{-\topsep}

Agents are trained in a goal-directed manner. Specifically, the agent is assigned random target values for velocity and lane change frequency that constitute its goal for that given episode. Each episode terminates when the agent either exhausts its time budget or its reward consistently falls below a predefined threshold. Upon completion of the training phase, the policy network weights are frozen and then used directly to govern the agent's driving behavior in subsequent simulations.

\subsection{Time-slice Mechanism Empowered Simulator Execution}
Previous research \cite{WANG2021102016,9200739} employed a multi-threaded Virtual Machine (VM) approach for simulation entities, where each request spawned separate threads responsible for maintaining states and emitting events. However, this approach has posed several significant challenges: \textbf{a) Hardware Dependency:} Simulations were tightly bound to the hardware configuration of the host system, making it difficult to reproduce and compare results across different hardware environments. \textbf{b) Impact of Non-Real-Time OS Scheduling:} Simulations become inaccurate due to the scheduling behavior of non-real-time operating systems. This scheduling idiosyncrasy could disrupt the expected sequencing and synchronization of operations and result in less precise experiment results. \textbf{c) Interference Effects under Computational Load:} In scenarios with concurrently executing multiple threads that exceeded the capacity of the system, interference between threads becomes significant as they compete for resources.

\begin{figure}[!h]
    \centering

    \begin{subfigure}{\textwidth}
        \centering
        \includegraphics[width=0.85\textwidth]{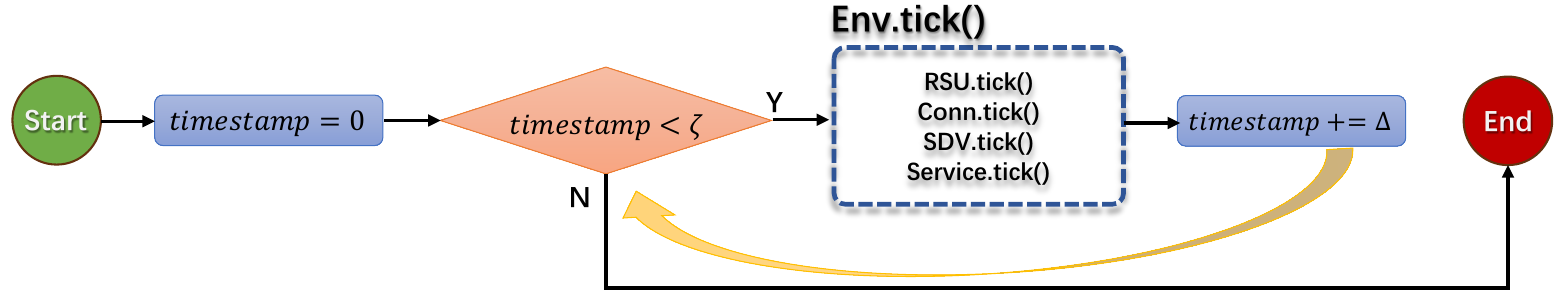}
        \caption{}
    \end{subfigure}

    \begin{subfigure}{0.49\textwidth}
        \centering
        \includegraphics[width=0.9\textwidth]{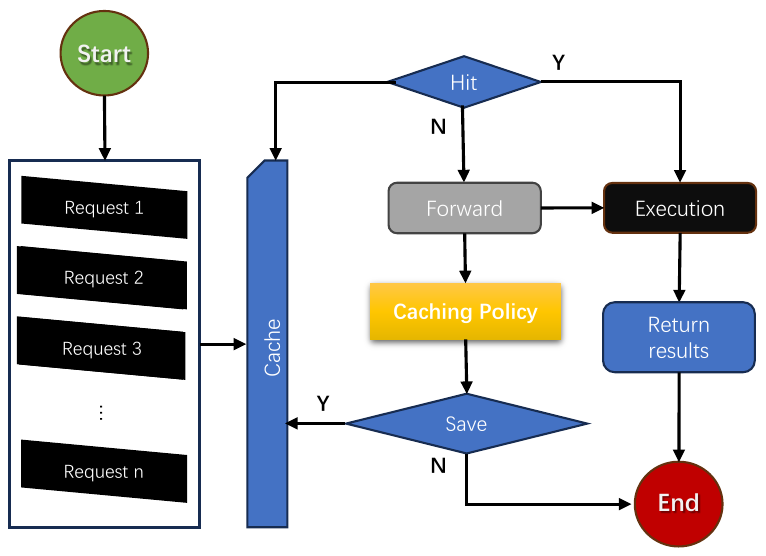}
        \caption{}
    \end{subfigure}
    \hfill
    \begin{subfigure}{0.49\textwidth}
        \centering
        \includegraphics[width=0.87\textwidth]{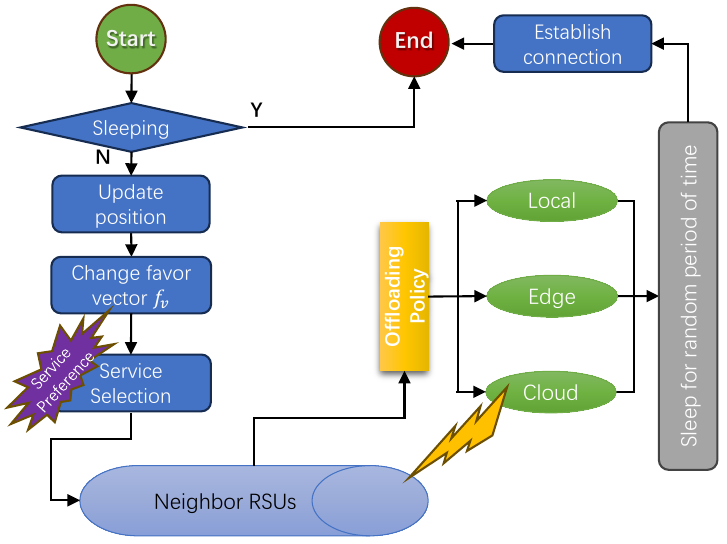}
        \caption{}
    \end{subfigure}
    
    \vspace{1em}
    
    \begin{subfigure}{0.49\textwidth}
        \centering
        \includegraphics[width=0.7\textwidth]{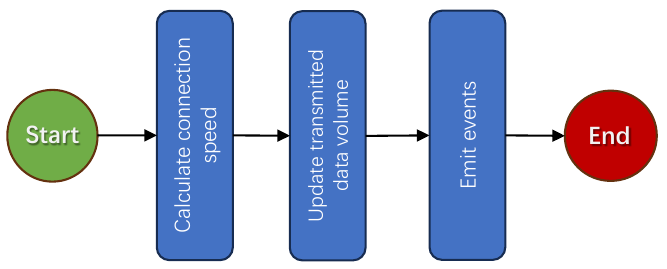}
        \caption{}
    \end{subfigure}
    \hfill
    \begin{subfigure}{0.49\textwidth}
        \centering
        \includegraphics[width=0.55\textwidth]{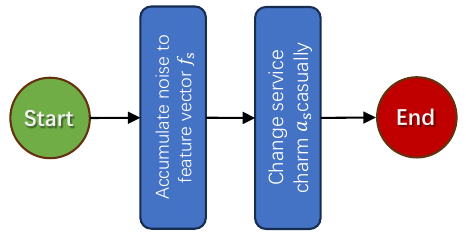}
        \caption{}
    \end{subfigure}
    
    \caption{Time-slice based execution model and entity update procedures in VEC-Sim. (a) The main simulation loop. (b)-(e) are the flowcharts of $tick()$ function for $RSU$, $SDV$, $Conn$, and $Service$ entities, respectively.}
    \label{FIG:6}
\end{figure}

To overcome the drawbacks of thread-based simulation, VEC-Sim incorporates a time-slice based execution model to enable more stable, reproducible and configurable simulations. Specifically, within each  time-slice, all entities (e.g., SDVs, Services, Conns) execute a $tick()$ function to refresh their status. This allows simulating the progression of the vehicular environment in a discrete, synchronous fashion. During the $tick()$, entities update their individual states and interact with the global environment. Figure \ref{FIG:6} presents a comprehensive workflow diagram encapsulating the key procedural steps employed throughout the entity status update process.

Figure \ref{FIG:6}(a) illustrates the main loop of VEC-Sim, which is responsible for repeatedly invoking the $env.tick()$ function and maintaining the time stamp. The $env.tick()$ function encapsulates time-slice update for all entities within the environment, calling the $tick()$ operations for all RSU, Conn, SDV, and Service objects. As depicted in Figure \ref{FIG:6}(b), for all RSUs in the system, each tick operation involves retrieving requests from the queue, checking for cache hits, forwarding requests in case of cache misses, applying user-defined cache policies to determine whether to cache the service, and finally offloading and executing the service. For SDVs, as shown in Figure \ref{FIG:6}(c), each $tick()$ encompasses actions such as position updates, interest vector modifications, service request selection, and random sleep periods. Regarding Conn objects (Figure \ref{FIG:6}(d)), the time-slice simulation process for these objects involves calculating the connection speed under current network conditions, updating transmission progress, and calling event functions. Figure \ref{FIG:6}(e) demonstrates the $service.tick()$ process, where each invocation of this function induces changes in the interest vector, thereby simulating the evolution of user preferences over time. This mechanism allows for a dynamic representation of shifting user interests within the simulated environment. The entire system updates through time-slice polling, enabling the simulation of the dynamic evolution of different entity states.

VEC-Sim's synchronized approach is a more controlled and elegant design. The time-slice based approach improves stability and reproducibility by providing a deterministic framework for simulations. Since all entities update their states within well-defined time intervals, the simulation results become more consistent and can be easily reproduced. This is particularly important for analyzing and evaluating various scenarios in the vehicular environment. However, this can potentially lead to performance issues, especially when dealing with a large number of entities. To address this concern, we incorporate the concept of Stepping that allows VEC-Sim to skip certain calculations in specific $\Delta$ time steps, rather than executing them in every time step. This optimization alleviates computational load and ensures smoother simulation execution.

\section{Experiments}
The experiments were running on a server equipped with Intel Core-i7 12700K CPU, 64GB RAM, and Nvidia RTX 4090 GPU. Simulations were executed on the Python 3.11.3 interpreter within Ubuntu 22.04.2 environment to ensure consistency across experimental trials. In this section, we first define the evaluation metrics used to assess the performance of the policies being simulated. Subsequently, experiments are presented to demonstrate VEC-Sim's capabilities in modeling real-world behavior, reproducing previous works findings, and conducting insightful case studies.

\subsection{Evaluation Metrics}
To quantitatively evaluate the performance of different resource scheduling strategies, VEC-Sim incorporates the following built-in metrics:

\vspace{-\topsep}
\begin{enumerate}[(a)]

    \item \textbf{Hit Rate:} The percentage of requests that can be served directly from the cache. This metric plays a pivotal role in reducing latency and improving the overall performance of the system, which can be calculated as
    \begin{equation}
        {hit\_rate}(r_i) = \frac{{\sum_{{req_{{v_i,s_k} }^n \in \{reqs \to r_i\}}} \mathbbm{l}_{{e_i.has\_cache(s_i)}}}}{{count\left( {req_{{v_i,s_k} }^n \in \{reqs \to r_i\}} \right)}} ,
        \label{EQ:HIT_RATE}
    \end{equation}
    where $\mathbbm{l}_{{e_i.has\_cache(s_i)}}$ is an indicator function that returns 1 if RSU $r_i$ has service $s_i$ cached, and 0 otherwise.

    \item \textbf{Average Response Time:} The average time elapsed from when a request is initiated to when the response is received, which can be formulated as
    \begin{equation}
        \overline{T}(r_i) = \frac{{\sum_{{req_{v_i,s_k}^n \in \{reqs \to r_i\}}} T(req)}}{{count\left( {req_{{v_i,s_k} }^n \in \{reqs \to r_i\}} \right)}} .
         \label{EQ:AVG_RESPONSE_TIME}
    \end{equation}

    \item \textbf{QoS:} QoS is a measure of the overall performance of the system, taking into account factors such as latency and throughput. VEC-Sim calculates QoS using
    \begin{equation}
        {QoS}(r_i) = \frac{1}{{P{(r_i)}^2}} \ln \left(1+{tr}_{effective}(r_i)\right) =  \frac{1}{{P{(r_i)}^2}} \ln \left(1 + \frac{{\sum_{{{req} \in \{reqs \to r_i\}}} {size}({req})}}{{\sum_{{{req} \in \{reqs \to r_i\}}} T(req)}}\right) , 
         \label{EQ:QoS}
    \end{equation}

    where QoS is proportional to effective transmission rate and inversely proportional to the square of power statistics $P$, which calls for an optimal policy that can minimize energy consumption and maximize ${tr}_{effective}$.

    \item \textbf{Service Offloading Load Balancing:} This indicator measures the evenness of load distribution during the offloading process. Load balancing factor $\mathcal{L}$ of the entire VEC network is expressed as
    \begin{equation}
         \mathcal{L} = \frac{1}{{\sqrt{\frac{{\sum_{i=1}^P \left[w(r_i) - \overline{w}\right]^2}}{P}}}} ,
          \label{EQ:LOAD_BALANCING}
    \end{equation}
    
    where $w(r_i)$ is the computational load on RSU $r_i$, and $\overline{w}$ is the average load across all RSUs. A lower value of $\mathcal{L}$ indicates more balanced load distribution.

\end{enumerate}

\subsection{Demonstration of Service Preference Mechanism}

In this section, we present a demonstration of how SDVs learn to discover and consume preferred services tailored to their individual preferences through the use of ratings. Our experimental setup involves 1000 services categorized into 5 clusters and 10 SDVs each features a 128-dim interest vector.

\begin{figure}[!h]
    \centering
    
    \begin{subfigure}{0.34\textwidth}
        \centering
        \raisebox{10mm}{
        \includegraphics[width=\textwidth]{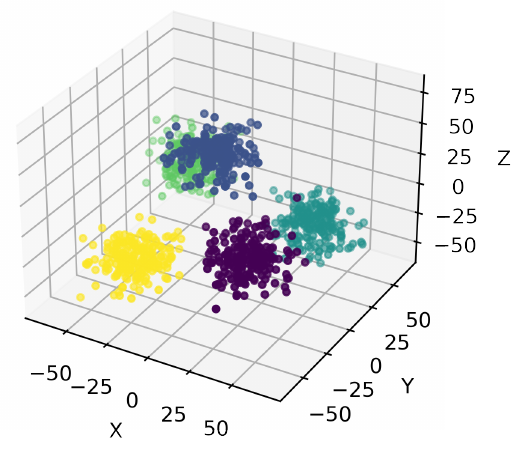}
        }
        \caption{}
    \end{subfigure}
    \hfill
    \begin{subfigure}{0.65\textwidth}
        \centering
        \begin{subfigure}{\textwidth}
            \centering
            \includegraphics[width=\textwidth]{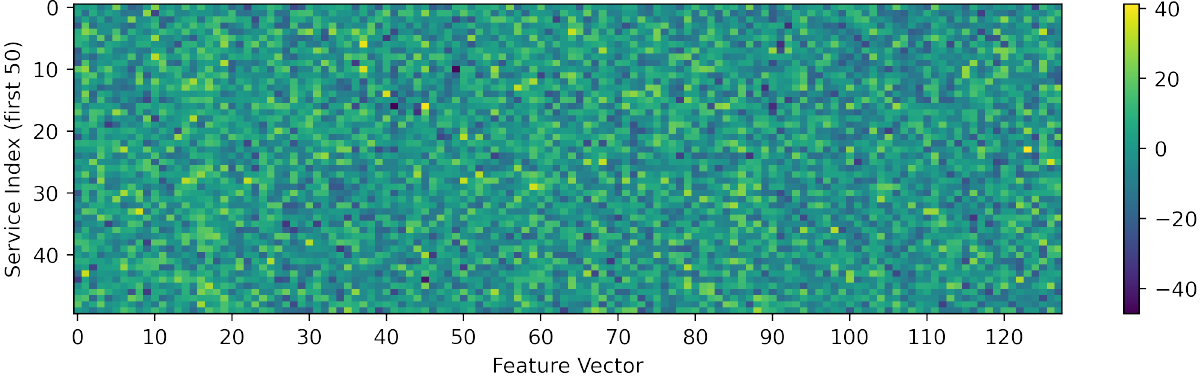}
            \caption{}
        \end{subfigure}
        
        \vspace{0.5em}
        
        \begin{subfigure}{\textwidth}
            \centering
            \includegraphics[width=\textwidth]{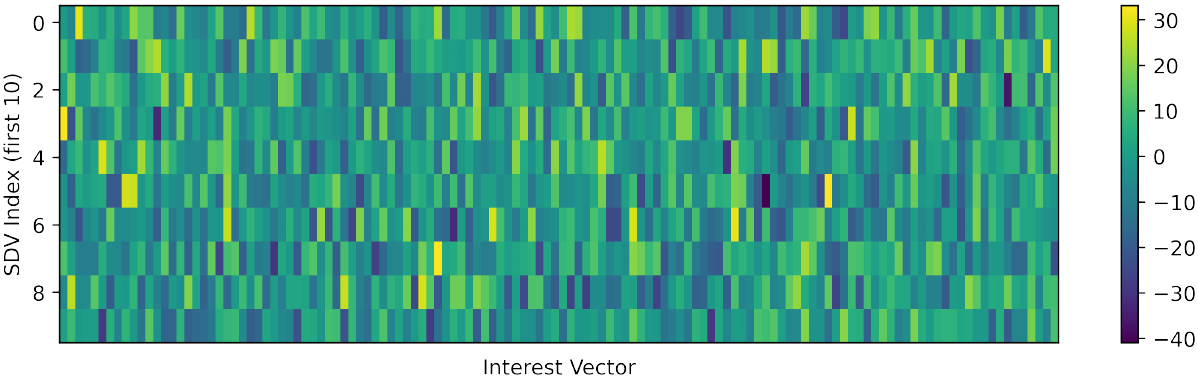}
            \caption{}
        \end{subfigure}

    \end{subfigure}
    
	\caption{Visualization of vector distribution. (a) PCA-based feature vector visualization. (b) Service feature vector heatmap. (c) SDV interest vector heatmap.}
	\label{FIG:7}
\end{figure}

To initiate our analysis comprehensively, we first verify the distribution of generated vectors. Figure \ref{FIG:7} illustrates the distributions of the generated service feature vectors and SDV preference vectors. In Figure \ref{FIG:7}(a), a 3D Principal Component Analysis (PCA) projection of the service feature vectors is presented, effectively highlighting the existence of five distinct clusters within the generated data. These clusters represent significant groupings of services with similar attributes, facilitating the understanding of service categorization. Figure \ref{FIG:7}(b) and Figure \ref{FIG:7}(c) are heatmaps of the service feature and SDV preference vectors, respectively. These heatmaps reveal the presence of Gaussian noise in the vectors. This observation is essential as it underscores the inherent variability and complexity within the user preference and service feature data.

\begin{figure}[!h]
	\centering
	\includegraphics[width=0.85\linewidth]{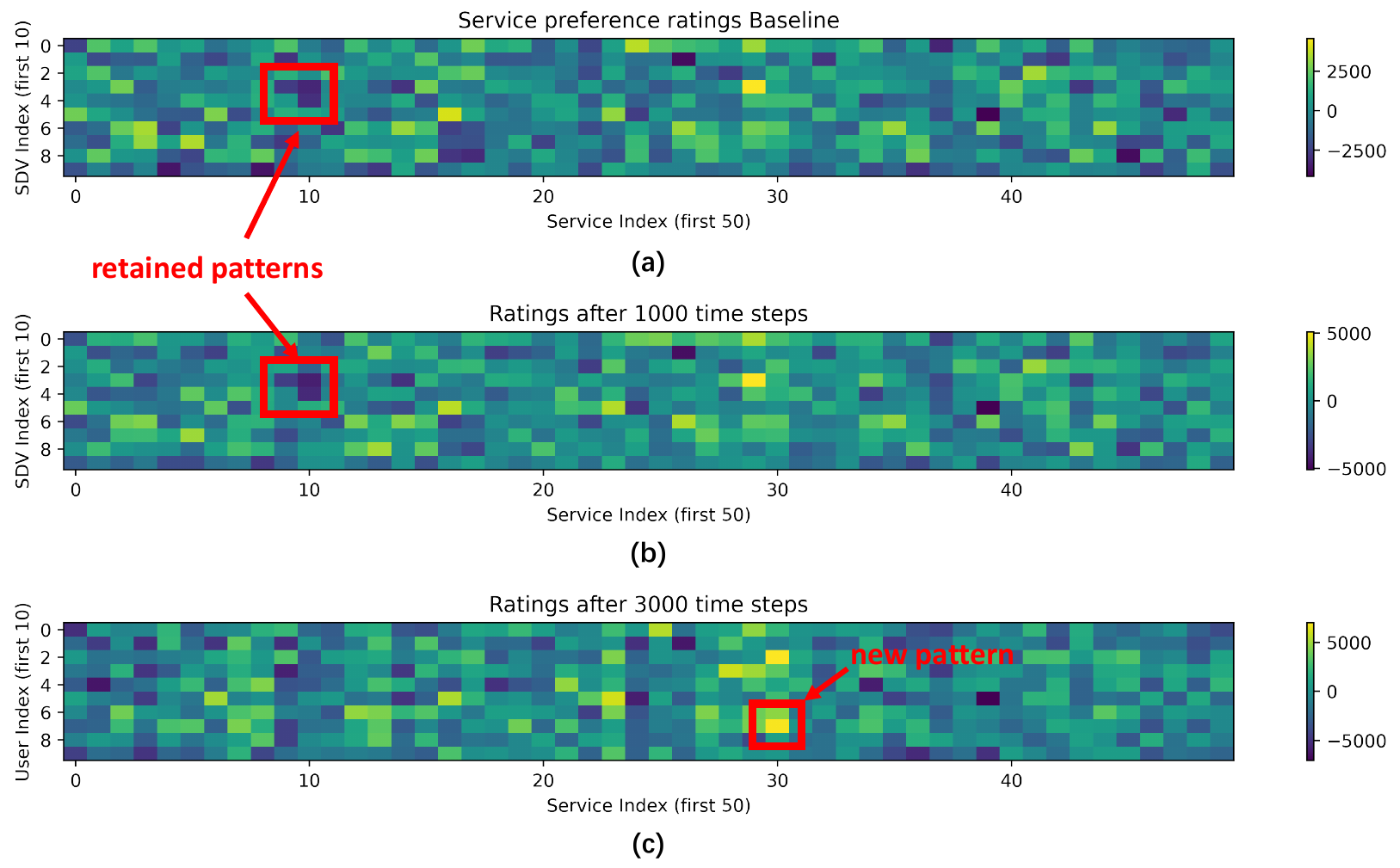}
	\caption{Evolution of service ratings over time. (a) Service preference ratings baseline. (b) Preference ratings after 1000 timesteps. (c) Preference ratings after 3000 timesteps.}
	\label{FIG:8}
\end{figure}

Figure \ref{FIG:8} illustrates the temporal evolution of service ratings as assessed by a representative SDV agent. These ratings reflect the agent's dynamically computed affinity for each service, adapting to its evolving preferences over time. Brighter colors in the heatmap denote larger values, signifying higher appeal. As discussed in Section \ref{section_4.3}, factors like marketing campaigns or emerging trends can lead to shifts in user preferences. To simulate the continuous evolution of user preferences, VEC-Sim plays a role by slightly adjusting vectors, resulting in corresponding changes in ratings. Figure \ref{FIG:8}(a) depicts the initial ratings, which are primarily determined by the initialization state of vectors. After 1000 timesteps, as shown in Figure \ref{FIG:8}(b), subtle alterations become evident in various regions of the heatmap. These changes indicate that the SDV agent's preferences have undergone some shifts. However, the overall structure largely retains its core preferences, suggesting that the agent's fundamental preferences remain relatively stable. Figure \ref{FIG:8}(c), after 3000 timesteps, illustrates the emergence of significantly more new patterns. This indicates that the SDV agent has further shifted its interests and even developed new preferences over the extended period of time.

\subsection{SDV Mobility Model Convergence Analysis}

\begin{figure}[!h]
	\centering

        \begin{subfigure}{\textwidth}
            \centering
            \includegraphics[width=0.7\textwidth]{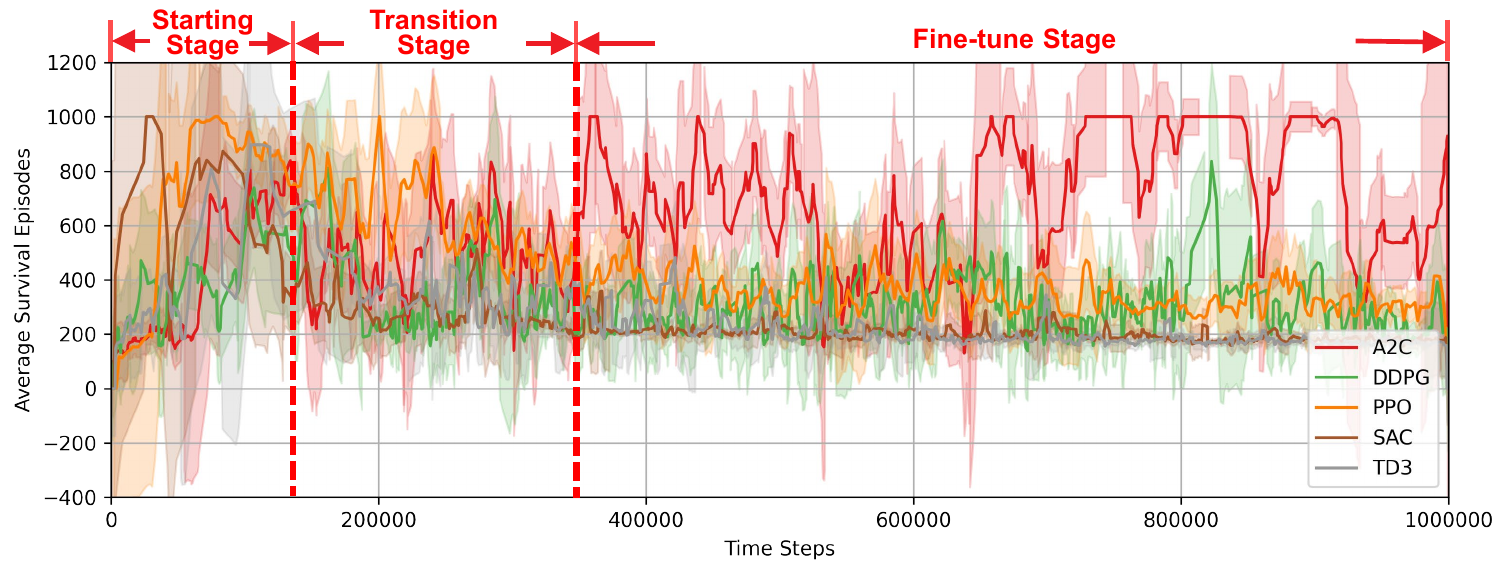}
            \caption{}
        \end{subfigure}
    
        \begin{subfigure}{\textwidth}
            \centering
            \includegraphics[width=0.7\textwidth]{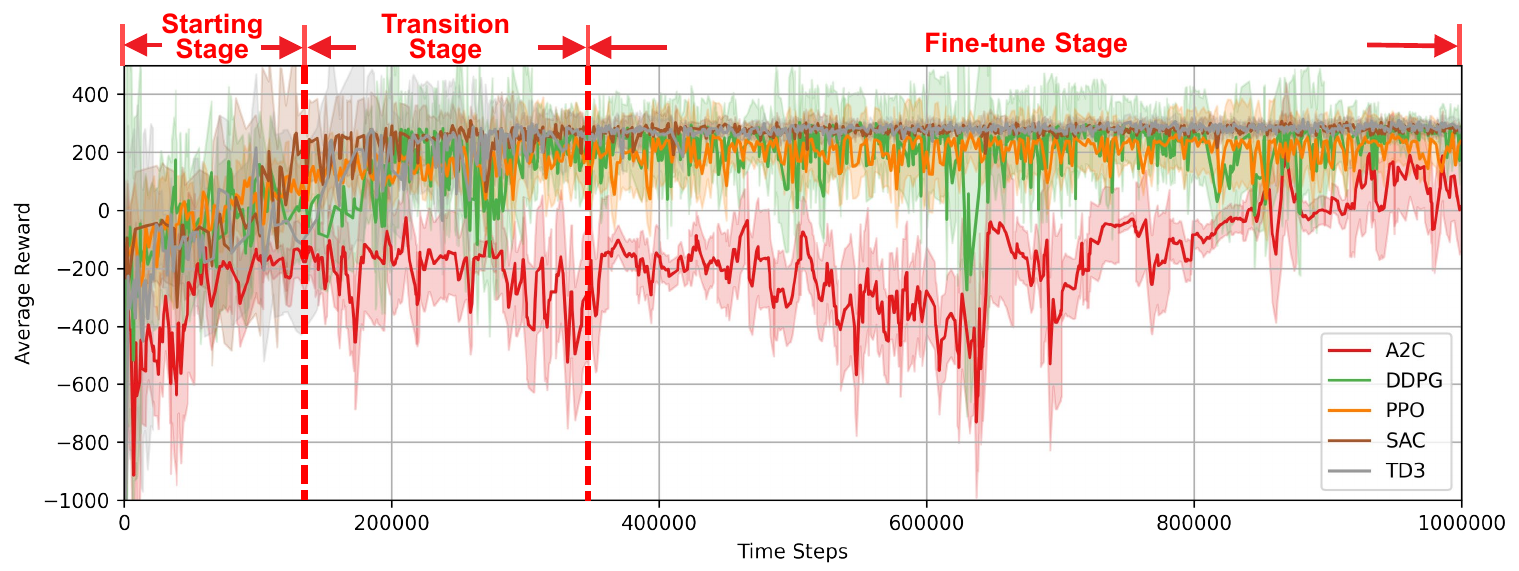}
            \caption{}
        \end{subfigure}

	\caption{Convergence analysis diagram of SDV mobility model under hyperparameter setting: batch size of 256, learning rate of $5 \times 10^{-4}$, discount factor of 0.99, replay buffer size of $10^6$ and epsilon factor from 0.95 to 0.1 with $10^{-4}$ decay per episode. (a) Number of surviving episodes over time, where lower values indicate more frequent environment resets due to rewards falling below the threshold, while higher values suggest more episodes reaching the time budget limit. (b) Average reward over time.}
	\label{FIG:9}
\end{figure}

In section \ref{section_4.4}, we presented the design of our SDV mobility model, which utilizes DRL agents for learning policies from the environment. To assess the feasibility of our proposed DRL-based mobility modeling approach and identify the most suitable DRL algorithm(s) that exhibit desirable convergence characteristics, it is essential to verify the convergence of our model. We selected a set of DRL algorithms well-suited for handling continuous action spaces, namely Asynchronous Advantage Actor-Critic (A3C), Deep Deterministic Policy Gradients (DDPG), Proximal Policy Optimization (PPO), Twin Delayed Deep Deterministic Policy Gradients (TD3), and Soft Actor-Critic (SAC). The convergence analysis diagram of the SDV mobility model is illustrated in Figure \ref{FIG:9}.

The learning process for each algorithm in Figure \ref{FIG:9} can be divided into three stages: \textbf{a) Starting Stage:} During this stage, both the average reward and surviving episodes show rapid growth. This can be attributed to the random initialization of policy network parameters. Thus, exploration is crucial as the agents have limited knowledge of the environment and need to gain initial experience \cite{9580706}. \textbf{b) Transition Stage:} The growth rate of average reward becomes less dramatic during this stage and may even experience a temporary dip. This is often accompanied by a decrease in the number of surviving episodes, indicating that the algorithms are breaking from local optima and transitioning to better policies. These observations signify agents' shift from random exploration to exploitation, as they start to refine their strategies. \textbf{c) Fine-tune Stage:} Most curves exhibit subtle fluctuations and tend to stabilize in this stage. At this point, the agents have largely concluded the exploration of new policies and shift their focus towards leveraging the existing strategies. The learning process reaches a state of convergence, where further improvements focus on fine-tuning the established policies rather than exploring entirely new ones.

Additionally, it is observed that average surviving episodes significantly decreases around 159k time steps for DDPG, PPO, SAC and TD3. This is due to the initial behavior of the agents, where they intentionally prolong their survival time (as defined by a threshold of 1000 episodes) by adopting strategies such as slowing down or rotating in place to maximize cumulative rewards. As the learning process advances, these initial behaviors are discarded in favor of more effective strategies, resulting in a decrease in the surviving episodes.

Regarding the comparison of algorithms, TD3 and SAC exhibit higher stability compared to DDPG and PPO. These two algorithms not only achieve higher rewards during training, but also exhibit lower variance. Specifically, TD3 and SAC show minimal fluctuations after 200k time steps, indicating an exceptional level convergence quality. It is evident that agents have learned the optimal strategies in a general context. The A2C algorithm displays a relatively slower convergence rate and introduces notable noise into the training process. While the precise underlying cause remains unknown, it is likely attributed to nuances in hyperparameter tuning and the algorithm's inherent design. However, it is worth noting that all algorithms eventually converge to similar values despite the slower convergence speed. The analysis confirms the convergence of the SDV mobility model. 

\subsection{Conducting Case Studies with VEC-Sim}

VEC-Sim's API design has been meticulously structured around an event-driven paradigm, offering researchers a versatile and elegant framework for customizing this platform. Besides, the modular API encapsulation also allows for seamless integration of bespoke resource scheduling strategies.

In this section, we perform case studies in the simulator, including reproduction of classical algorithms, analysis of geographical distribution impact, etc. These studies are conducted in comparison with real-world ground truths, thereby providing an additional layer of validation, enhancing our confidence in the simulator's capabilities. Simulations are carried out within a 10 km $\times$ 10 km scope, spanning a simulation duration of 10,000 timesteps with stepping set to 10. If not specified otherwise, the default simulation parameters are utilized for consistency throughout this section, as outlined in Table \ref{tb2}.

\begin{table}[width=0.7\linewidth,cols=2,pos=h]
\caption{Simulation Parameters}\label{tb2}
\begin{tabular*}{0.7\linewidth}{@{}p{0.35\linewidth}p{0.35\linewidth}@{}}
\toprule
\textbf{Parameter} & \textbf{Value}  \\ 
\midrule
    Number of SDVs $H$ & 1000 \\
    SDV computation capacity & 10 GFLOPS \\
    SDV target speed & 40km/h \\
    SDV cache size &  $\{4, 8, 16\}$GB randomly \\
    Number of RSUs $P$ & 20 \\
    RSU coverage range $g_i$ & 500m $ \sim $ 3km (Normal Distribution) \\
    RSU computation capacity & 100 GFLOPS \\
    Number of services $K$ & 10e6 \\
    Service size distribution & 1MB $ \sim $ 1GB (Zipf) \\
    Service computation distribution & 1 GFLOPS \textasciitilde{} 10 TFLOPS (Zipf) \\
    Service feature vector clusters & 5 \\
    V2R channel bandwidth $\mathbb{B}$ & 500Mbps \\
\bottomrule
\end{tabular*}
\end{table}

\subsubsection{Benchmark of Classical Caching Algorithms}
In order to further evaluate the feasibility and observe the behavior of VEC-Sim, we conducted a comparative analysis of five well-established caching algorithms \cite{HASSLINGER2023109583}. The baselines selected for benchmark are as follows:

\vspace{-\topsep}
\begin{itemize}
\setlength{\itemsep}{0pt}
\setlength{\parsep}{0pt}
\setlength{\parskip}{0pt}
    \item \textbf{Random Replacement:} A straightforward approach that evicts a random service from the cache when it's full, requiring minimal bookkeeping.

    \item \textbf{First In First Out (FIFO):} Evicts services based on their entry order, with the earliest items added to the cache being replaced first.

    \item \textbf{Least Recently Used (LRU):} It keeps tracking of access recency and prioritizes the eviction of the least recently accessed service.

    \item \textbf{Least Frequently Used (LFU):} Calculates the number of accesses within a specific time frame to determine cache actions. When the cache is full, it prioritizes the eviction of the least accessed services.

    \item \textbf{CLOCK:} A variant of the FIFO algorithm that aims to approximate LRU with lower overhead. It keeps a circular buffer and a reference pointer, replacing the item where the pointer lands.
\end{itemize}
\vspace{-\topsep}

Experiments of different cache sizes are conducted for each algorithm to align with real-world scenarios. To obtain the necessary data, we utilized the interfaces provided by the statistics object within VEC-Sim. By setting up anchor points within the simulation environment, performance data is collected at predefined time intervals. The simulation results are presented in Table \ref{tb3}.

\begin{table}[cols=6,pos=h]
\caption{Performance comparison of caching policies under different cache sizes}\label{tb3}
\begin{threeparttable}
\begin{tabular}{>{\centering}m{0.1\linewidth}>{\centering}m{0.1\linewidth}>{\centering}m{0.15\linewidth}>{\centering}m{0.22\linewidth}>{\centering}m{0.1\linewidth}>{\centering\arraybackslash}m{0.15\linewidth}} 
\hline
\textbf{Cache Size} & \textbf{Algorithm} & \textbf{Hit Rate (\%), \\Eq. (\ref{EQ:HIT_RATE})} & \textbf{Avg Response Time (s), \\Eq. (\ref{EQ:AVG_RESPONSE_TIME})} & \textbf{QoS\tnote{a)}, \\Eq. (\ref{EQ:QoS})} & \textbf{Space Utilization\tnote{b)} (\%)} \\ 
\hline
\multirow{5}{*}{4GB} & Random & 14.12 & 9.83 & 0.18 & 85.47 \\
 & FIFO & 30.79 & 8.61 & 0.40 & 85.41 \\
 & LRU & 44.25 & 6.09 & 1.04 & 85.53 \\
 & LFU & \textbf{56.33} & \textbf{5.07} & \textbf{1.26} & \textbf{89.34} \\
 & CLOCK & 46.51 & 6.15 & 1.01 & 88.63 \\ 
\hline
\multirow{5}{*}{8GB} & Random & 21.34 & 8.72 & 0.37 & 94.80 \\
 & FIFO & 47.06 & 7.51 & 0.83 & \textbf{97.64} \\
 & LRU & 63.90 & 4.63 & 1.24 & 94.15 \\
 & LFU & \textbf{70.45} & \textbf{3.89} & \textbf{1.53} & 90.23 \\
 & CLOCK & 66.53 & 4.13 & 1.44 & 89.27 \\ 
\hline
\multirow{5}{*}{16GB} & Random & 39.56 & 5.19 & 0.95 & 97.65 \\
 & FIFO & 78.12 & 4.54 & 1.57 & 99.47 \\
 & LRU & 84.69 & 3.40 & 1.93 & 95.89 \\
 & LFU & \textbf{92.36} & \textbf{2.36} & \textbf{2.04} & 98.40 \\
 & CLOCK & 81.43 & 3.52 & 1.93 & \textbf{99.76} \\
\hline
\end{tabular}
    \begin{tablenotes}
     \item[a)] The QoS parameter ranges from 0 to $+\infty$. It incorporates power consumption and effective transmission rate to quantify the system's overall service quality, with higher values indicating better performance.
     \item[b)] The proportion of cached content in the system relative to the total cache capacity. This metric reflects the efficiency of cache resource utilization and the degree of storage space usage.
   \end{tablenotes}
   \end{threeparttable}
\end{table}

As cache capacity expanded from 4GB to 16GB, all algorithms demonstrated substantial performance enhancements, consistent with the theoretical expectation that increased cache size allows for greater data retention, thus improving hit rates. LFU consistently outperformed its counterparts, particularly at larger cache capacities, achieving a peak hit rate of 92.36\% and a minimum average response time of 2.36s at 16GB cache size. This observation underscores the simulator's accurate representation of hot-spot services and temporal access patterns prevalent in VEC environments, where LFU's strategy of prioritizing frequently accessed items proves highly effective in scenarios with distinct service popularity disparities. The performance gap between sophisticated algorithms (LFU, LRU) and their simpler counterparts (Random, FIFO) widened as cache size increased, with LFU surpassing Random by 133.47\% in hit rate at 16GB cache size. This growing disparity illustrates VEC-Sim's capacity to replicate complex real-world scenarios, where the benefits of intelligent caching strategies become more pronounced with increased resources, mirroring behaviors observed in actual VEC systems. As anticipated, no significant variations in space utilization were observed among the algorithms, with utilization ranging from 95.89\% to 99.76\% at 16GB cache size. This observation aligns with the absence of specific optimizations for load balancing and fragmentation in classic caching strategies, further validating VEC-Sim's accurate modeling of these algorithms' behaviors in realistic VEC environments.

The observations in Table \ref{tb3} collectively affirm VEC-Sim's capability to consistently reproduce key characteristics of real-world VEC scenarios, including cache size impacts, hot-spot behaviors, and temporal access patterns, establishing it as a robust platform for VEC research. Notably, VEC-Sim exhibited consistent operation throughout extended simulations with no errors and unexpected behaviors, further demonstrating its stability during prolonged operation.

\subsubsection{Impact of Edge Server Placement on VEC Network Performance}

\begin{figure}[!h]
    \centering
     
    \begin{subfigure}{0.5\textwidth}
        \centering
        \includegraphics[width=\textwidth]{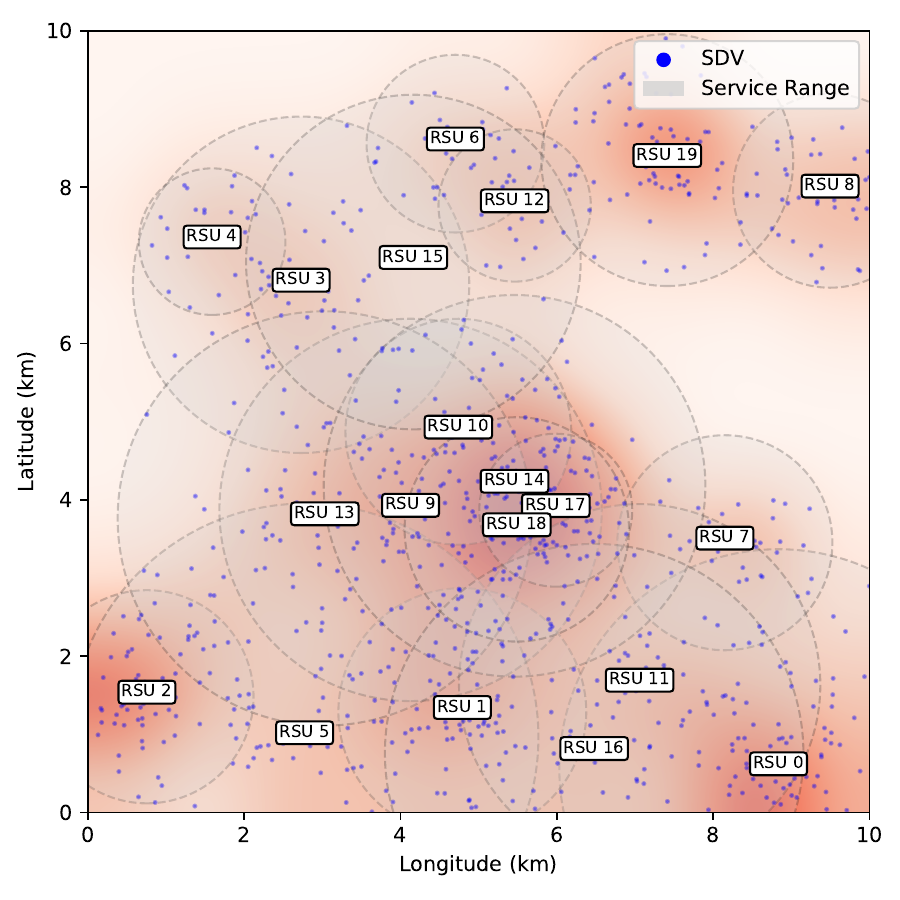}
        \caption{}
    \end{subfigure}
    \hfill
    \begin{subfigure}{0.49\textwidth}
        \centering
        \begin{subfigure}{\textwidth}
            \centering
            \includegraphics[width=\textwidth]{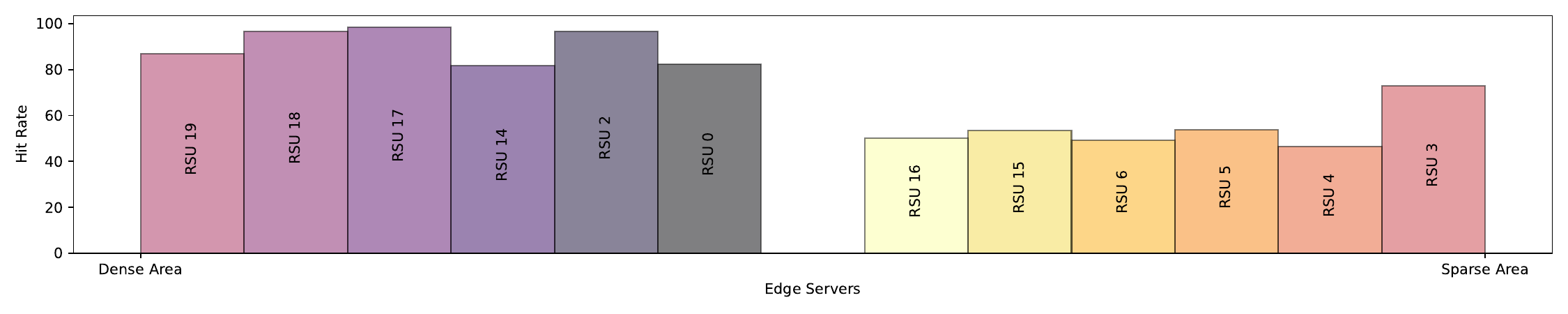}
            \caption{}
        \end{subfigure}
        
        \begin{subfigure}{\textwidth}
            \centering
            \includegraphics[width=\textwidth]{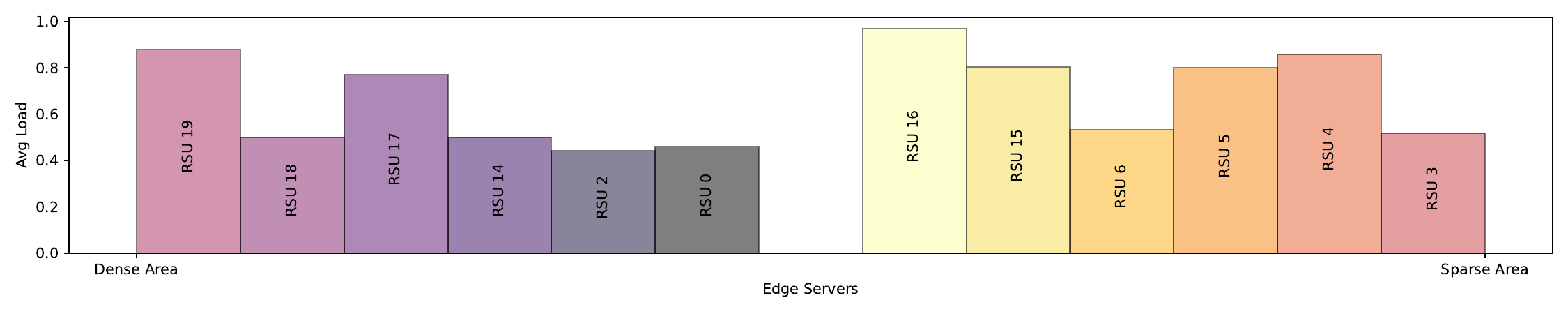}
            \caption{}
        \end{subfigure}
        
        \begin{subfigure}{\textwidth}
            \centering
            \includegraphics[width=\textwidth]{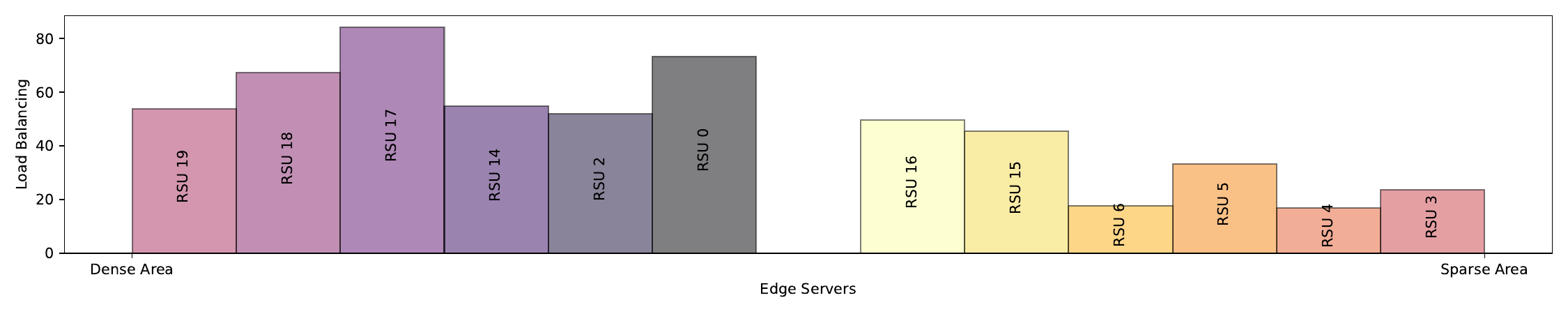}
            \caption{}
        \end{subfigure}
        
        \begin{subfigure}{\textwidth}
            \centering
            \includegraphics[width=\textwidth]{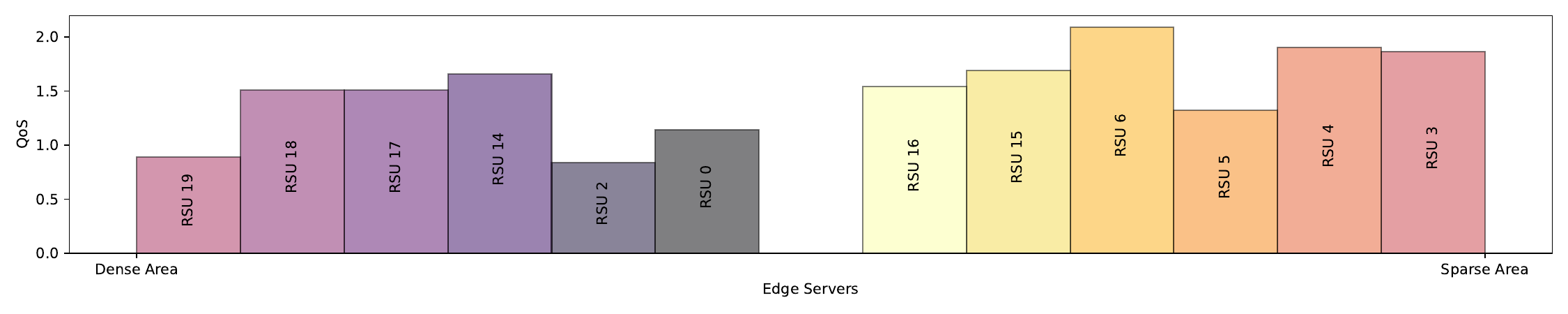}
            \caption{}
        \end{subfigure}
    \end{subfigure}

	\caption{Impact of RSU and SDV geographical distribution on VEC network performance. (a) Heatmap showing the geographical distribution of SDVs and RSUs, with color intensity representing SDV density. Subsequently, we compare the performance in sparse and dense areas based on (b) Hit Rate, (c) Average Load, (d) Load Balancing, and (e) QoS.}
	\label{FIG:10}
\end{figure}

In addition to replicating existing work, VEC-Sim is a valuable tool for conducting hypothetical explorations, enabling researchers and decision-makers to make validation and identify risks in uncharted territories. For instance, the placement of RSUs has always been a challenging task \cite{10167729} in real-world scenarios. This involves striking a delicate balance between effective coverage, redundancy, and cost-effectiveness. Considerations such as user density should be factored in to reduce the distance between users and RSUs, while hardware and maintenance costs should also be taken into account to ensure economic feasibility. In our experiment, we employ VEC-Sim to investigate the impact of deployment density on VEC network performance. We adopt the offloading policy proposed by \cite{9284036}, and the results are presented in Figure \ref{FIG:10}.

The placement of RSUs as depicted in Figure \ref{FIG:10}(a) exhibits an uneven distribution. Notably, RSUs 14, 15, and 17 are positioned closely together, surrounded by a cluster of SDVs, whereas RSU 4 and 5 have fewer SDVs in their vicinity. Analysis of Figure \ref{FIG:10}(b) reveals that denser deployments of RSUs result in higher hit rates for the surrounding SDVs, with dense areas averaging 90.46\% compared to 54.28\% in sparse areas. This substantial difference is likely attributed to the inter-RSU collaboration mechanism stated in section \ref{section_3.4}. The difference in average load between sparse and dense deployments is not significant in Figure \ref{FIG:10}(c). This suggests that although dense regions experience higher demand, the workload is distributed among a larger number of RSUs. On the other hand, sparse areas have lower request volumes due to fewer RSUs, resulting in a balancing effect between the two scenarios. This finding implies that the overall workload of placement in Figure \ref{FIG:10}(a) is distributed reasonably well across both sparse and dense regions. Furthermore, Figure \ref{FIG:10}(d) reveals a significant enhancement in load balancing within dense areas. Figure \ref{FIG:10}(e) unveils a remarkable phenomenon: RSUs 14, 17, and 18, situated in dense areas, surpass the performance of some RSUs in sparse regions (e.g., RSUs 5 and 16). Notably, RSU 14 achieves a QoS of 1.66, while RSU 5 in the sparse area only attains 1.323. This observation provides compelling evidence for the efficacy of the selected service offloading strategy \cite{9284036} under evaluation. Aligning with the authors' assertions, the strategy not only accomplishes workload balancing but also ensures superior QoS in high-population-density areas. Dense areas maintain an average QoS of 1.258, marginally lower than the 1.736 observed in sparse areas.

These findings further underscore the capabilities of VEC-Sim, as the practical implementation and validation of the algorithm corroborate the original authors' results. The simulator's proficiency in replicating real-world scenarios and generating accurate results positions it as a valuable tool for researchers in the field.

\section{Conclusion and Future Work}
In this paper, we introduced VEC-Sim, a simulator specifically tailored for vehicular networking scenarios. VEC-Sim boasts a modular design, capable of accommodating both caching and offloading simulation-based functionalities. To augment the realism of VEC-Sim, we incorporated essential features such as service preference modeling, favor shift mechanisms, service upload simulations, and time-slice modeling. Subsequently, we conducted a series of comprehensive experiments and case studies to validate the feasibility and effectiveness of our proposed solution.

However, there are still several areas that warrant further investigation. An imperative direction for future research is the integration of native support for Vehicle-to-Vehicle (V2V) communication and collaboration among multiple RSUs. While our initial approach simplified these aspects, addressing them comprehensively holds immense potential for enhancing system capabilities. Furthermore, it's essential to to leverage techniques such as multi-threading and coroutines to fully utilize the available computing resources. Ultimately, conducting a broader spectrum of case studies and real-world experiments for comparative analysis will yield deeper insights into the system's behavior across various scenarios.



\printcredits

\section*{Declaration of Competing Interest}
The authors declare that they have no known competing financial interests or personal relationships that could have appeared to influence the work reported in this paper.

\section*{Data availability}
Data will be made available on request.

\section*{Acknowledgement}
This work was supported by the National Natural Science Foundation of China under Grant (No. 62372242 and 92267104), and in part by Natural Science Foundation of Jiangsu Province of China under Grant (No. BK20211284).

\bibliographystyle{unsrt}
\bibliography{cas-refs}


\bio{}
\textbf{Fan Wu} is currently a postgraduate student at the School of Software, Nanjing University of Information Science and Technology, China. His research interests include mobile edge computing, fault diagnosis, etc.
\endbio


\bio{}
\textbf{Xiaolong Xu} received the Ph.D. degree in computer science and technology from Nanjing University, China, in 2016. He was a Research Scholar with Michigan State University, USA,from April 2017 to May 2018. He is currently a Full Professor with the School of Software, Nanjing University of Information Science and Technology. He has published more than 100 peer-review articles in international journals and conferences, including the IEEE TPDS, IEEE T-ITS, IEEE TII, ACM TOSN, ACM TOMM, ACM TIST, IEEE ICWS, ICSOC, etc. He was selected as the Highly Cited Researcher of Clarivate 2021 and 2022. His research interests include Big data, Internet of Things and Deep learning.
\endbio


\bio{}
\textbf{Muhammad Bilal} received the Ph.D. degree from the Korea University of Science and Technology. He served as a Postdoctoral Research Fellow at Korea University’s Smart Quantum Communication Center. In 2018, he joined Hankuk University of Foreign Studies, South Korea, where he was as an Associate Professor with the Division of Computer and Electronic Systems Engineering. In 2023, he joined Lancaster University, where he is working as a Senior Lecturer (Associate Professor) with the School of Computing and Communications. With over 14 years of experience, Dr. Bilal has actively participated in numerous research projects funded by prestigious organizations like IITP, MOTIE, NRF, and NSFC, etc. He is the author/co-author of 140+ articles published in renowned journals and holds multiple US and Korean patents. His research interests include Network Optimization, Cyber Security, the Internet of Things, Vehicular Networks, Information-Centric Networking, Digital Twin, Artificial Intelligence, and Cloud/Fog Computing.
\endbio


\bio{}
\textbf{Xiangwei Wang} is currently a postgraduate student at the School of Software, Nanjing University of information Science and Technology, China. His research interests include cloud-edge collaboration and computer vision.
\endbio


\bio{}
\textbf{Hao Cheng} is currently a postgraduate student at the School of Software, Nanjing University of information Science and Technology, China. His research interests include edge computing and federated learning.
\endbio


\bio{}
\textbf{Siyu Wu} is currently a postgraduate student at the School of Software, Nanjing University of Information Science and Technology, China. His research interests include IoV, recommendation systems, etc. 
\endbio

\end{document}